\documentclass[preprint,12pt]{elsarticle}

\usepackage{amsmath}
\usepackage{float}
\usepackage{comment}
\usepackage{amssymb}
\usepackage{color}
\usepackage[left]{lineno}
\usepackage{nameref}
\usepackage{amstext}
\usepackage{textcomp}

\newcommand*{\xmax}{$X_\text{max}$}
\newcommand*{\mxmax}{X_\text{max}}
\newcommand*{\gcm}{\text{g}/\text{cm}^2}

\newcommand*{\zmax}{z_\text{max}^\text{GDAS}}
\newcommand*{\fig}{{\sc Fig.}}
\newcommand*{\figs}{{\sc Figs.}}
\newcommand*{\eq}{{\sc Eq.}}
\newcommand*{\eqs}{{\sc Eqs.}}
\newcommand*{\tab}{{\sc Tab.}}
\newcommand*{\ngdasgd}{N^\text{\tiny{GDAS}}_\text{\tiny{GD}}}
\newcommand*{\nusgd}{N^\text{\tiny{US}}_\text{\tiny{GD}}}
\newcommand*{\ngdashf}{N^\text{\tiny{GDAS}}_\text{\tiny{HF}}}
\newcommand*{\ngdashfdry}{N^\text{\tiny{GDAS}}_\text{\tiny{HF,dry}}}

\begin{document}
\begin{frontmatter}

\title{Computing the electric field from Extensive Air Showers using a realistic description of the atmosphere}
\author[suba]{F. Gat\'e\corref{cau}}
\author[suba,usn]{B. Revenu\corref{cau}}
\author[suba]{D. Garc\'{\i}a-Fern\'{a}ndez}
\author[home]{V. Marin}
\author[suba,usn]{R. Dallier}
\author[suba]{A. Escudi\'e}
\author[suba,usn]{L. Martin}

\address[suba]{Subatech, IMT Atlantique, CNRS, Universit{\'e} de Nantes, France}
\address[usn]{Station de Radioastronomie de Nan\c{c}ay, Observatoire de Paris, PSL Research University, CNRS, Universit\'e d'Orl\'eans, Nan\c{c}ay, France}
\address[home]{Nantes, France}
\cortext[cau]{florian.gate@lapp.in2p3.fr, revenu@in2p3.fr}

\begin{abstract}
The composition of ultra-high energy cosmic rays is still poorly known and constitutes a very important topic in the field of high-energy astrophysics. Detection of ultra-high energy cosmic rays is carried out via the extensive air showers they create after interacting with the atmosphere constituents.
The secondary electrons and positrons within the showers emit a detectable 
electric field in the kHz-GHz range.
It is possible to use this radio signal for the estimation of the atmospheric depth of maximal development of the showers \xmax, with a good accuracy and a duty cycle close to $100\%$. This value of \xmax\ is strongly correlated to the nature of the primary cosmic ray that initiated the shower. We show in this paper the importance of using a realistic atmospheric model in order to correct for systematic errors that can prevent a correct and unbiased estimation of~\xmax.

\end{abstract}


\end{frontmatter}

\date{\today}

\section{Introduction}
\label{sec:intro}

Recently a lot of efforts have been put into determining the mass composition of cosmic rays using the radio signal \cite{Bezyazeekov:2015ica,2016Natur.531...70B,gatearena2016}. Several methods exist by now with different approaches but the goal is the same: reconstructing the atmospheric depth of the shower maximum, \xmax, where the number of particles is maximum. This atmospheric depth is highly correlated to the mass of the primary cosmic ray. To be competitive, the uncertainty on its estimation should be close to or better than
that achieved with the fluorescence technique ($\sim 20~\gcm$, see~\cite{xmaxpao2010}). The composition of the highest-energy cosmic rays (above $1$~EeV) is still poorly known, since it is difficult to measure composition using a surface detector that only samples the shower at ground level. Besides, the fluorescence light technique, more apt for composition measurements, has a duty cycle of the order of $14$\%~\cite{fdpao2010}, making it difficult to provide \xmax\ measurements for a large number of showers at the highest energies. The radio technique, consisting in the measurement of the electric field induced by the extensive air showers created by cosmic rays, could be an excellent alternative to obtain the \xmax\ with an almost $100$\% duty cycle. Extracting the \xmax\ using the radio signal relies on an atmospheric model. The electric field emission is highly beamed towards the direction of propagation of the shower and the shape of its distribution at the ground level depends on the distance between the point of maximum emission and the shower core. This property can be exploited to reconstruct \xmax\ from the radio signal. However, to make this method accurate, one needs to know the atmospheric depth corresponding to a given distance with precision. The electric field measured by the antennas strongly depends on the characteristics of the atmosphere in which secondary shower particles evolve: air density, air refractive index at radio frequencies, temperature, pressure and humidity. For a long time, simulation codes computing this electric field assumed a standard atmosphere. Nowadays, with high precision measurements on large radio arrays running continuously such as AERA~\cite{glaserarena2016}, it has become important to refine this atmospheric model. Indeed, it is clear that the atmospheric characteristics vary significantly with time (day/night effect and seasonal variations) and these variations are responsible for systematic uncertainties that can prevent an accurate estimation of the \xmax. Ideally, we need to know the atmospheric state at the time a shower is detected. This is possible using the Global Data Assimilation System~\cite{gdas} (GDAS) data. In this paper, we show how we use these data together with a standard atmospheric model for the highest altitudes to compute an accurate air density model as a function of altitude at the time of the detection of the event. The knowledge of the air density and humidity ratio also allows to compute the realistic air refractive index which is needed for the amplitude and time structure of the signal.
Several descriptions of the atmosphere are in use in different simulation codes such as SELFAS~\cite{selfas2011}, ZHAireS~\cite{zhaires2012} and CoREAS~\cite{coreas2013}.
We show that the choice of the atmospheric model induces uncertainties in the atmospheric depths up to some tens of $\gcm$ which is comparable to the uncertainty on the \xmax\ obtained with the fluorescence data.
The paper is organized as follows. In section~\ref{sec:geo}, we briefly present the geometrical description of the shape of the Earth and its atmosphere and the atmospheric depths computations. In section~\ref{sec:PCasp} we describe the GDAS data and its use to build a realistic atmospheric model that we will use to calculate the atmospheric depths and the air refractive index. We compare the results with those obtained assuming the basic US Standard model \cite{Usstd}. In section~\ref{sec:influence} we quantify the influence of the air refractive index and air density profiles calculated with the GDAS data on the produced electric fields. Then, in section~\ref{sec:recoxmax} we study the case of a simulated shower which develops in the atmospheric conditions of a sample day. We show that using the US Standard model on the \xmax\ estimation leads to biased results, unless we use the same atmospheric conditions than those of the day and time of the detected (here simulated) event.
In this paper, we will note $\mathbf{V}$ the shower axis and $\mathbf{B}$ the geomagnetic field.

\section{Geometry of the atmosphere}
 \label{sec:geo}
\begin{figure}[H]
 \begin{center}
  \includegraphics[scale=0.2]{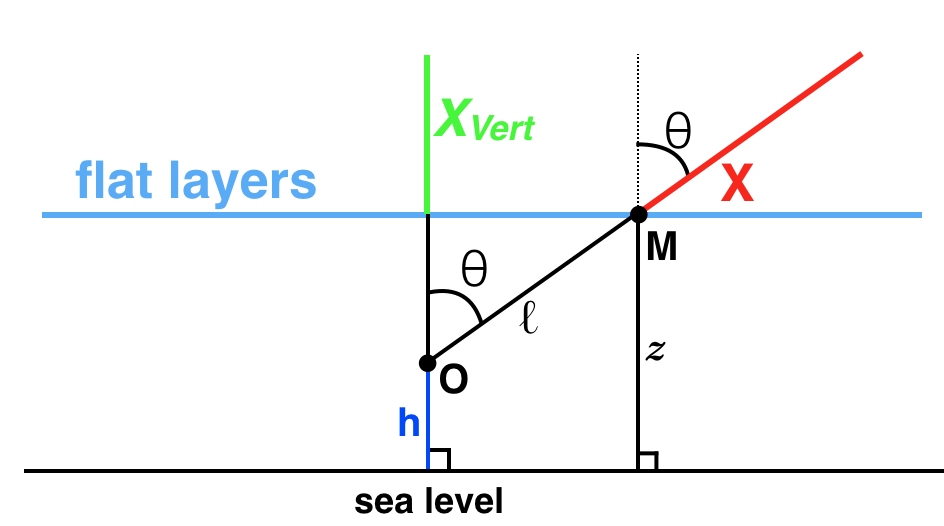}
   \includegraphics[scale=0.2]{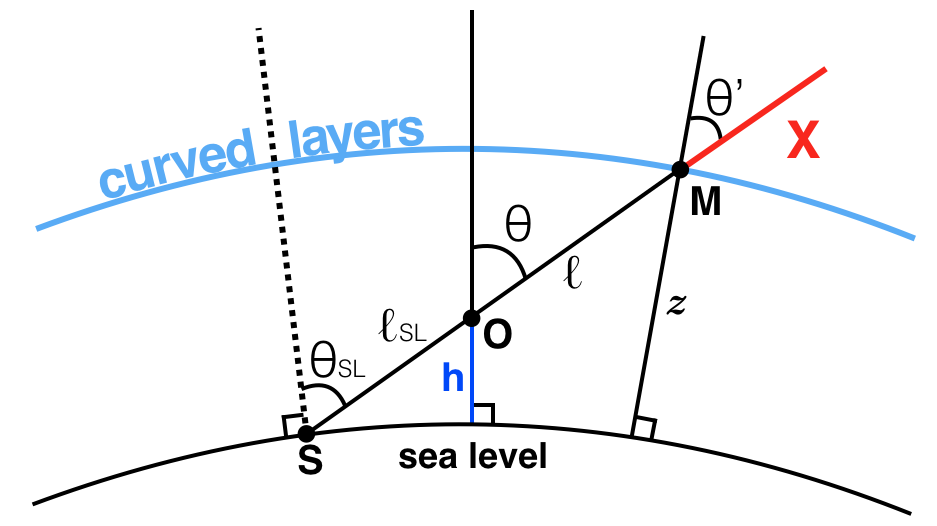}
   \caption{Left: flat atmosphere/Earth geometry. Right: spherical geometry.}
\label{fig:sketch}
  \end{center}
\end{figure}

Usually, the shape of the atmosphere is taken as flat or spherical. The spherical shape is taken into account when dealing with inclined showers, typically for zenith angles $\theta\geqslant 60^\circ$. In~\fig~\ref{fig:sketch}, we present both descriptions.
The atmospheric depth at distance $\ell$ from observer~$O$ and corresponding to an elementary path $\mathrm{d}\ell$ is given by $\mathrm{d}X_\text{slant}=  \rho(z(\ell)) \,\, \mathrm{d}\ell$, where $\rho$ is the air density and $z$ the altitude above sea level. In the flat approximation $\mathrm{d} z = \mathrm{d}\ell \cos\theta$ where $\theta$ is the zenith angle ---~between the vertical at $O$ and $(OM)$~--- so that $\mathrm{d}X_\text{slant} = \rho(z)\,\mathrm{d}z / \cos \theta =  \mathrm{d}X_\text{v} / \cos \theta$, where $\mathrm{d}X_\text{v}$ is the vertical elementary atmospheric depth. After integration we obtain:
\begin{equation}
X_{\text{slant}}(\ell) =  X_\text{v} (z(\ell)) / \cos \theta.\label{eqv}
\end{equation}
$X_\text{v} (z)$ represents the vertical atmospheric depth; it is known as the Linsley's parameterization when considering the US Standard model and provides the integrated atmospheric depth traversed vertically from "infinity" (i.e. where $\rho$ is negligible, before entering the atmosphere) to altitude~$z$.
The flat approximation is thus correct for vertical showers but considering the accuracy that radio methods intend to achieve, a comparison to a spherical description is necessary for inclined showers. The expression of the atmospheric depth in \eq~\ref{eqv} does not apply when~$\theta \ne 0$ because the atmospheric layers are curved. Moreover at a position $M$, the zenith angle $\theta^\prime$ is not the same than the angle $\theta$ at $O$ (see \fig~\ref{fig:sketch} right).
We consider an observer $O$ at the altitude $h$. The radius of the Earth is denoted~$R$. A point $M$ on the shower axis is located at an altitude $z$ (above the sea level). The zenith angle at $M$ depends on its position along the shower axis: it is $\theta$ for~$M = O$ (corresponding to an observer located at an altitude $h$). A simple geometrical calculation gives:

\begin{eqnarray}
\ell &=& \sqrt{(R+z)^2 - (R+h)^2 \sin^2 \theta } - (R+h) \cos \theta \nonumber \\
z &=& \sqrt{\ell^2 + (R+h)^2 + 2\ell(R+h) \cos \theta } - R \nonumber \\
\cos \theta' &=&  \sqrt{ 1-  \left( \frac{R+h}{R+z} \right)^2 \sin^2 \theta } \nonumber
\end{eqnarray}

\begin{figure}[H]
 \begin{center}
  \includegraphics[scale=0.28]{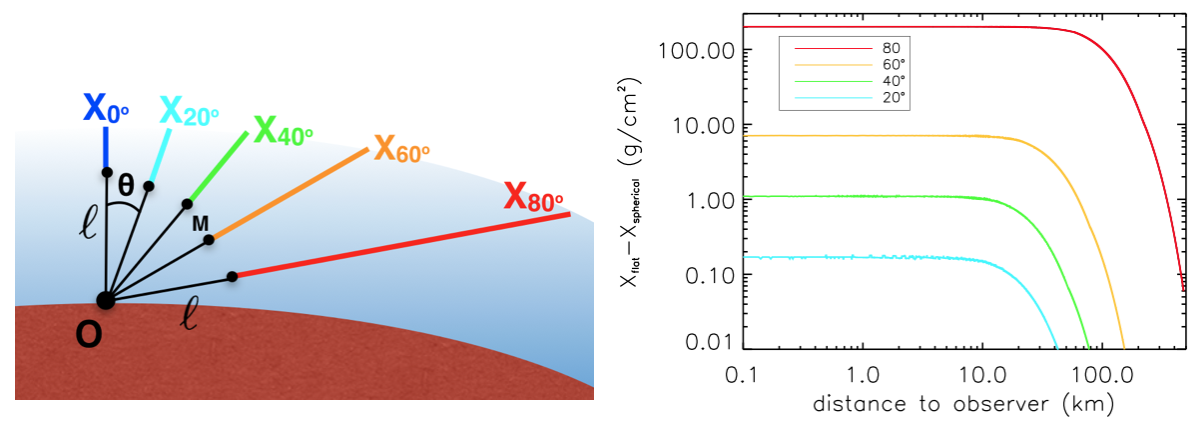}
   \caption{Right: differences in atmospheric depth obtained with the flat approximation (using \eq~\ref{eqv}) and the spherical description (using \eq~\ref{eqs}) for several zenith angles. The observer is located at the sea level and the shower hits the ground at the position of the observer. The distance-to-observer axis corresponds to the distance $\ell$ of \fig~\ref{fig:sketch} and is indicated in the left part of the figure.}
\label{fig:flat_curve}
  \end{center}
\end{figure}

The atmospheric slant depth is calculated numerically by integrating the atmosphere density along the shower axis:
\begin{equation}
X_{\text{slant}}(\ell) =  \int_{\ell}^{\infty} \rho(z(\ell')) \, \mathrm{d}\ell'\label{eqs}
\end{equation}
Where $\rho(z(\ell'))$ is the air density at a given altitude $z$ corresponding to a particle-to-observer distance $\ell'$ along the shower axis.
A comparison is made between the two descriptions in \fig~\ref{fig:flat_curve}: we choose an observer $O$ at sea level and a shower with a zenith angle $\theta$. The atmopheric depth crossed by the shower from outer space up to a distance $\ell$ to the observer along the axis is computed either with the flat approximation or the spherical description.
Both descriptions give equal results for a vertical shower ($\theta=0^\circ$).
Using the flat approximation leads to errors of the order of $10~\gcm$ for zenith angles larger than $60^\circ$. In the seek of accuracy, we should be very cautious with the flat approximation, even for not too inclined showers. In SELFAS, we always use the spherical description, independently of the zenith angle.

Apart from the atmospheric depths, we also checked the effect on the electric field computations. We found that one really needs to consider the spherical shape only for inclined showers ($\theta\geqslant 60^\circ$).

\section{Physico-chemical aspects of the atmosphere}
 \label{sec:PCasp}
 
The variations of the meteorological conditions are studied for the CODALEMA experiment. In the following sections, only data for the location of Nan\c{c}ay, France, are presented.
 
 \subsection{The GDAS data}
 \label{subsec:GDAS}

The characteristics of the atmosphere that are needed for computing the electric field emitted by air showers are the air refractive index ($\eta$) and density ($\rho$) at any altitude $z$. These parameters depend on relative humidity ($R_h$), temperature ($T$) and total pressure ($P$) that vary on a daily basis.

\begin{figure}[H]
 \begin{center}
 \includegraphics[scale=0.33]{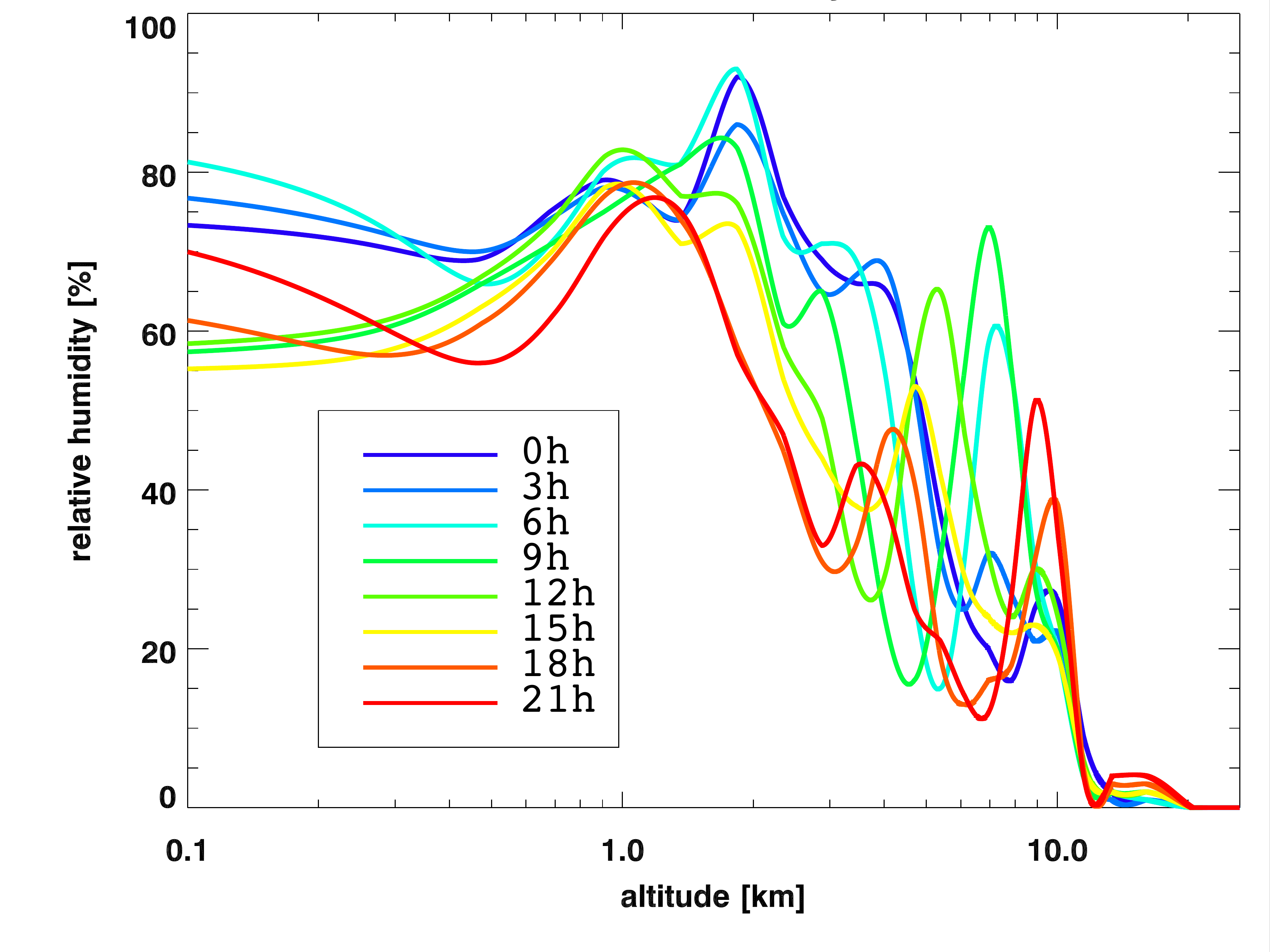}
   \caption{Daily variations of the relative humidity as a function of the altitude, using the GDAS data at Nan\c{c}ay on March 18, 2014.}
\label{fig:dRh}
  \end{center}
\end{figure}

As an illustration, we present in \fig~\ref{fig:dRh} the relative humidity as a function of the altitude from the GDAS data on March 18, 2014. We see that at a given altitude, the variations are very important according to the time of the day and consecutively, the same holds for the air density and index values.

\begin{figure}[H]
 \begin{center}
 \includegraphics[scale=0.3]{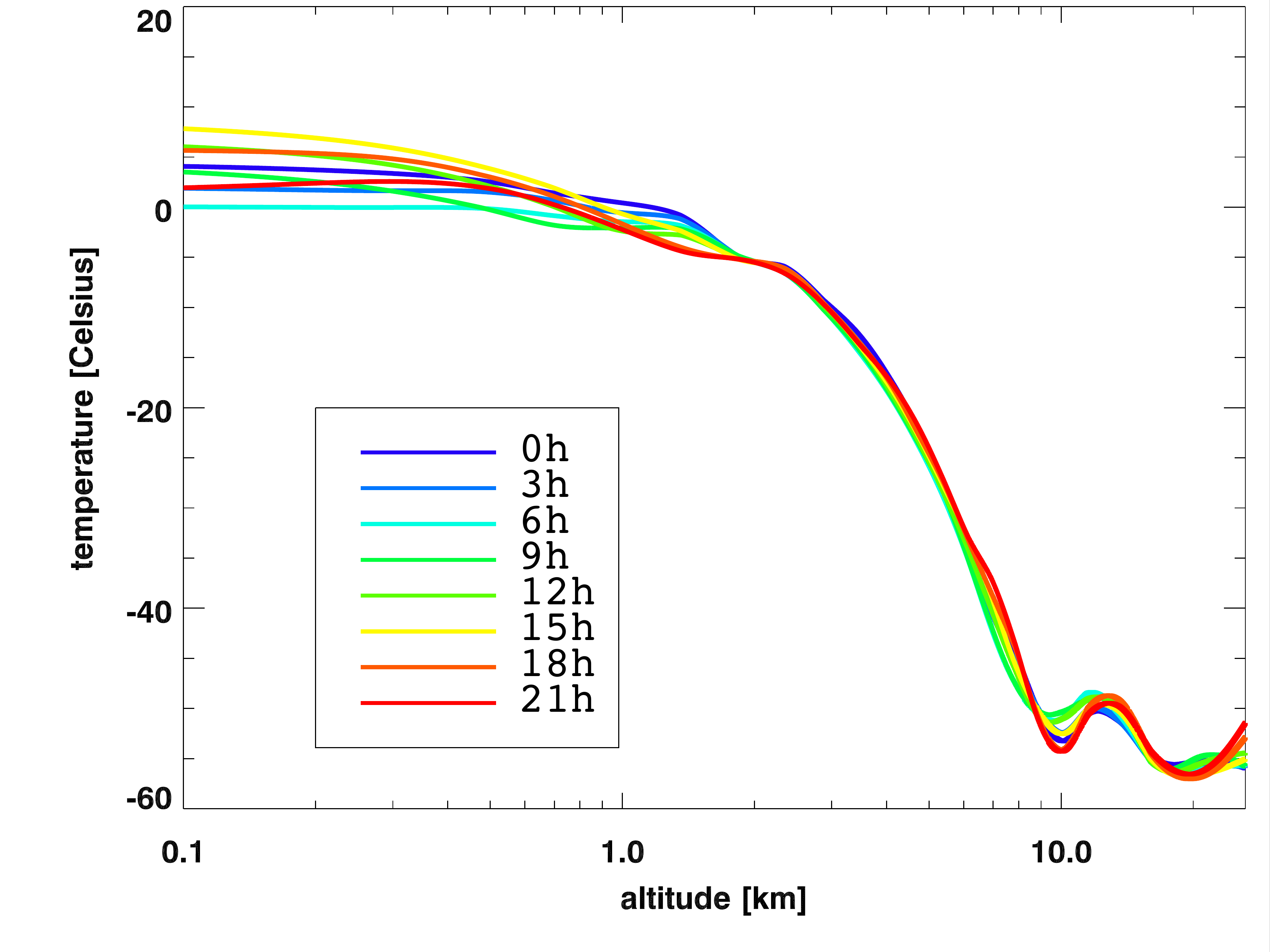}
 \includegraphics[scale=0.3]{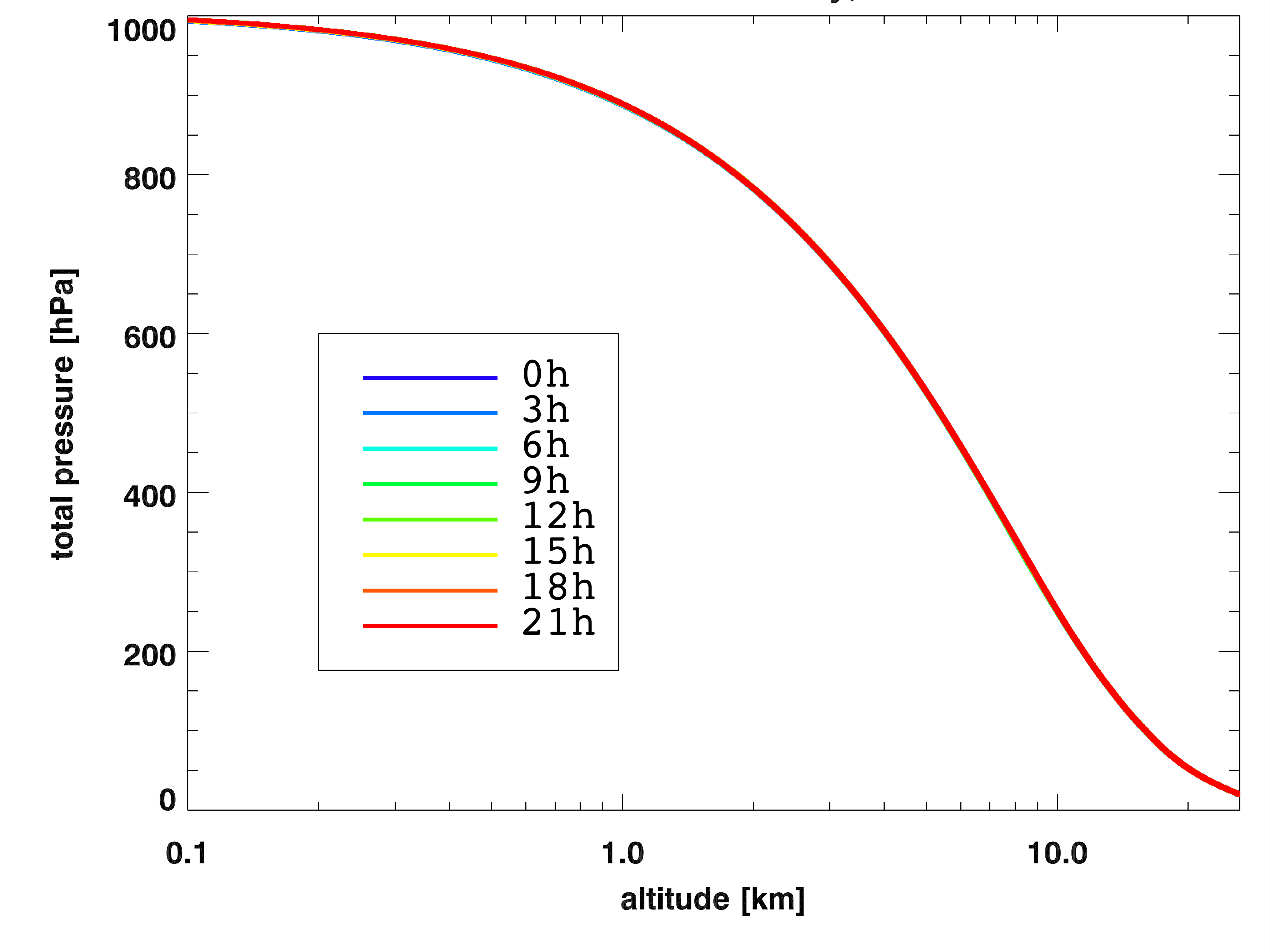}
   \caption{Daily variations of the temperature (top) and pressure (bottom) as a function of the altitude, using the GDAS data at Nan\c{c}ay on March 18, 2014.}
\label{fig:dT}
  \end{center}
\end{figure}

In \fig~\ref{fig:dT}, we show the same plot but for the temperature (top) and pressure (bottom). For temperature, above an altitude of $3-4$~km the variations are negligible as a function of time. The pressure is not varying significantly over time at fixed altitude and can also be taken as constant with time. However the latter quantities can vary more importantly over longer timescales. In this example of a single day, we can conclude that the precise knowledge of the pressure, temperature and relative humidity is mandatory in order to accurately compute the air index and density profiles. The values displayed in \figs~\ref{fig:dRh} and~\ref{fig:dT} were obtained from the GDAS which provides a database of measurements of physicochemical characteristics of the atmosphere. 

Each GDAS file contains a week of data and one must extract the ones corresponding to the desired location. The files contain measurements for every $3$ hours at the surface and $23$ geopotential heights up to an altitude of $\zmax=26$~km above sea level.\\

The results of the simulation of the EAS-induced electric field depend on the air index and density models of the atmosphere in which the shower develops. The adopted approach to provide SELFAS with realistic air profiles along with a proper geometrical description of the atmospheric layers from the GDAS data is explained in the next sections. Detailed comparisons between the US Standard model and  the GDAS profiles, as well as the consequences on the $X_{\text{max}}$ reconstruction will be presented.
Among all the available parameters provided by the GDAS, we use the pressure $P$ in~hPa, the geopotential height $Z_g$ in~gpm\footnote{geopotential meters}, the temperature $T$ in~K and the relative humidity $R_h$ in~\%.
As the GDAS provides data at given geopotential meters one must convert them into meters above sea level. The conversion formula is provided in the~\nameref{sec:Appendix}.

\subsection{Air density profile}
 \label{subsec:density}

\begin{figure}[!ht]
 \begin{center}
 \includegraphics[scale=0.2]{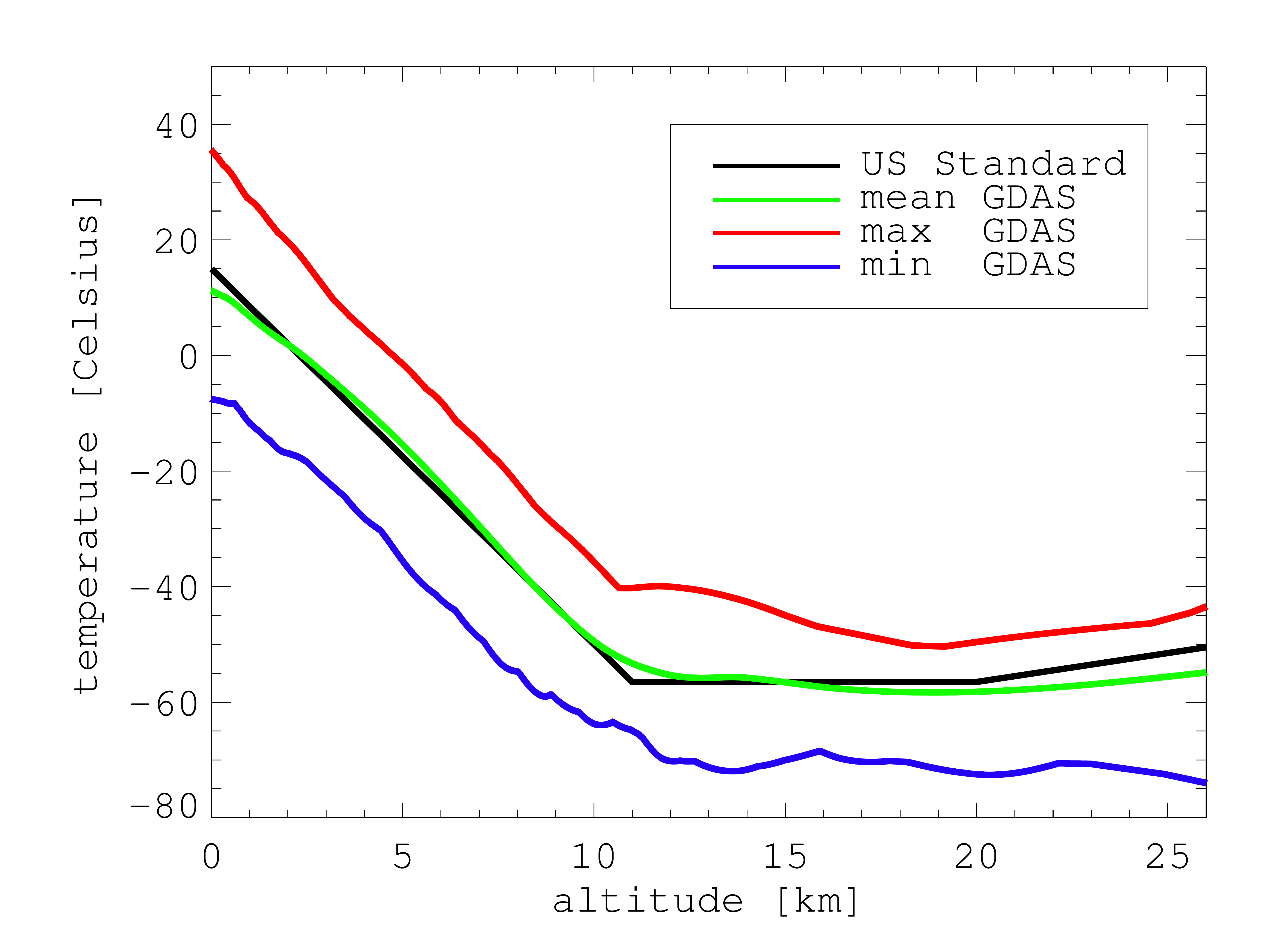}
 \caption{Temperature as a function of altitude, using the US Standard model (in black), the mean of the GDAS data for the year $2014$ (in green) and the minimum and maximum values of the GDAS data in blue and red respectively. At fixed altitude, the temperature can vary by $\pm 20^\circ$C with respect to the US Standard temperature.}
\label{fig:temp}
  \end{center}
\end{figure}

The air density as a function of the altitude is computed from the ideal gas law, taking into account the relative humidity:
\begin{equation}\label{egp}
\rho(z)=\frac{p_d(z(Z_g,\phi)) M_d + p_v(z(Z_g,\phi)) M_v}{R\, T(z(Z_g,\phi))},
\end{equation}
where $z(Z_g,\phi)$ is the altitude above sea level corresponding to the geopotential altitude~$Z_g$ at a latitude~$\phi$, $p_d$ and~$p_v$ are the partial pressures of dry air and water vapor,~$M_d$ and~$M_v$ are
the molar masses of dry air and water vapor,~$T$ (in~$K$) is the temperature and~$R$ is the universal gas constant. The formula used to calculate the saturation vapor pressure~$p_\text{sat}$ can be found in~\cite{buck2} and is a modification of a previous parameterization explained in~\cite{buck1}:
\begin{gather}
p_d=P-p_v \quad \text{with} \quad p_v=R_h\, p_\text{sat} \quad \text{and} \nonumber \\
p_\text{sat} = 6.1121 \exp\left[\left(18.678-\frac{T}{234.5}\right)\left(\frac{T}{257.14+T}\right)\right] \quad \text{(}T\text{ in $^\circ$C)}\label{eqpsat}
\end{gather}
This formula is accurate in the range~$[-80;+50]^\circ\text{C}$ which is suitable in our case if we refer to \fig~\ref{fig:temp} that shows the temperature profiles as a function of the altitude in Nan{\c c}ay for the year~$2014$. The minimum temperature during that year in the range $[0;\zmax]$~km is $-75^\circ$C and the maximum is $40^\circ$C.
We compare the temperature profiles from sea level to an altitude of $\zmax=26$~km as \eqs~\ref{egp} and~\ref{eqpsat} are only used up to this altitude. In SELFAS we need to know the air density profile up to an altitude of $100$~km, well above the GDAS limit because showers can start to develop well above $\zmax$. Between sea level and $\zmax$, we obtain the air density at any altitude by interpolation of the $23$ GDAS data points. Above $\zmax$, we use the US Standard air density profile as described in~\cite{Usstd} with a scaling factor ($f_B$) to ensure continuity with the GDAS data, calculated as: 

\begin{equation}
f_B = \frac{\rho_\text{GDAS}(\zmax)}{\rho_\text{USstd}(\zmax)}
\label{fb}
\end{equation}

The US Standard profile can be retrieved easily from~\cite{dutch} up to $100$~km of altitude but as a function of geopotential meters that one has to convert again in geometric altitude. After this procedure, the air density profile is known from sea level to an altitude of $100$~km:
\begin{equation}
\begin{split}
&\bullet \text{if}~z(\ell) > \zmax:  \quad \rho(z(\ell)) = \rho_\text{USstd}(z(\ell)) \times f_B \\
&\bullet \text{if}~z(\ell) < \zmax:  \quad \rho(z(\ell)) = \rho_\text{GDAS}(z(\ell))
\end{split}
\label{rho}
\end{equation}

 In order to estimate the seasonal and day/night systematics, a comparison is made for every possible GDAS profiles for the year $2014$ (i.e. one profile every $3$~hours along the year). 

\begin{figure}[H]
 \begin{center}
  \includegraphics[scale=0.3]{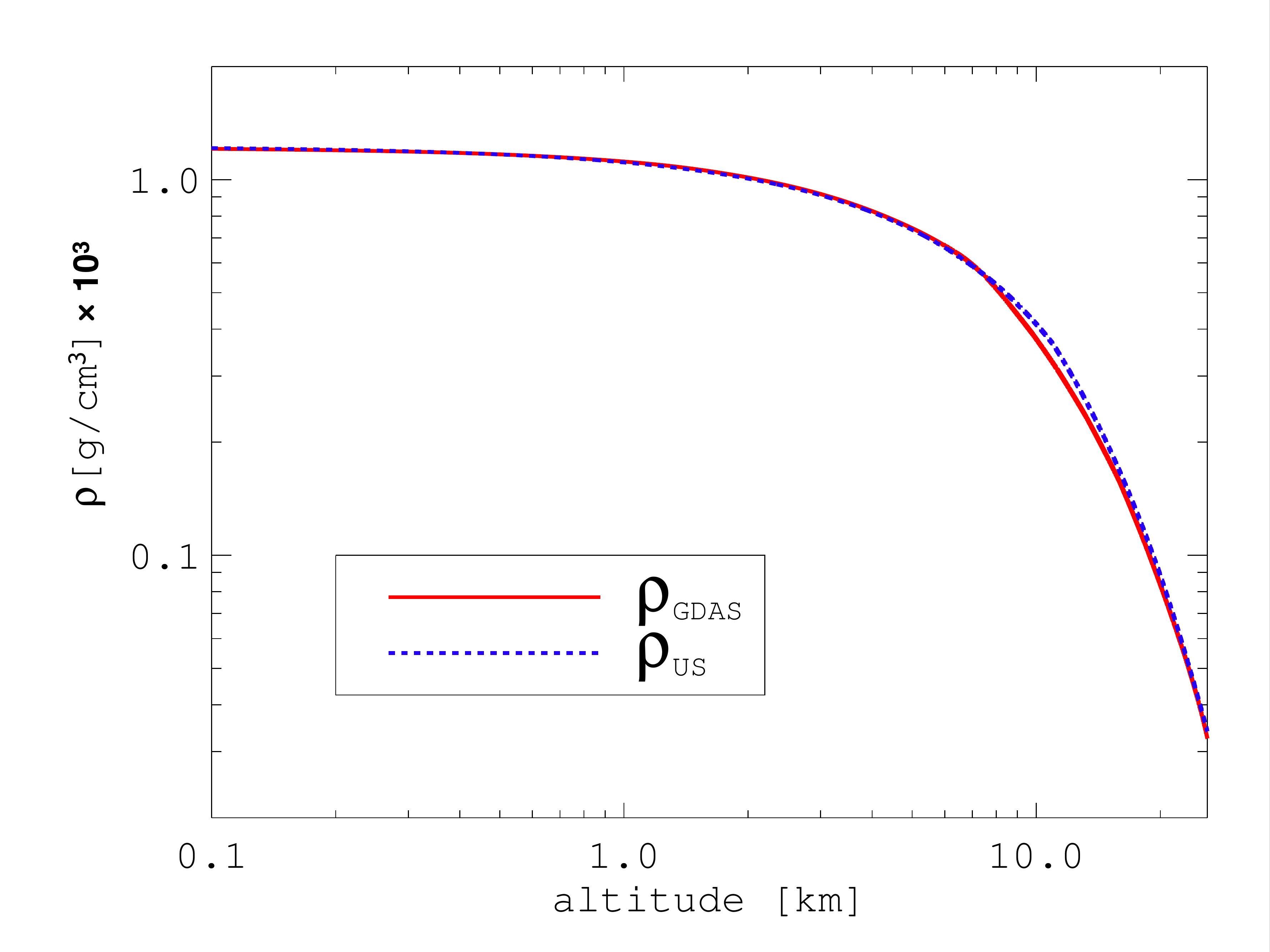}
 \includegraphics[scale=0.3]{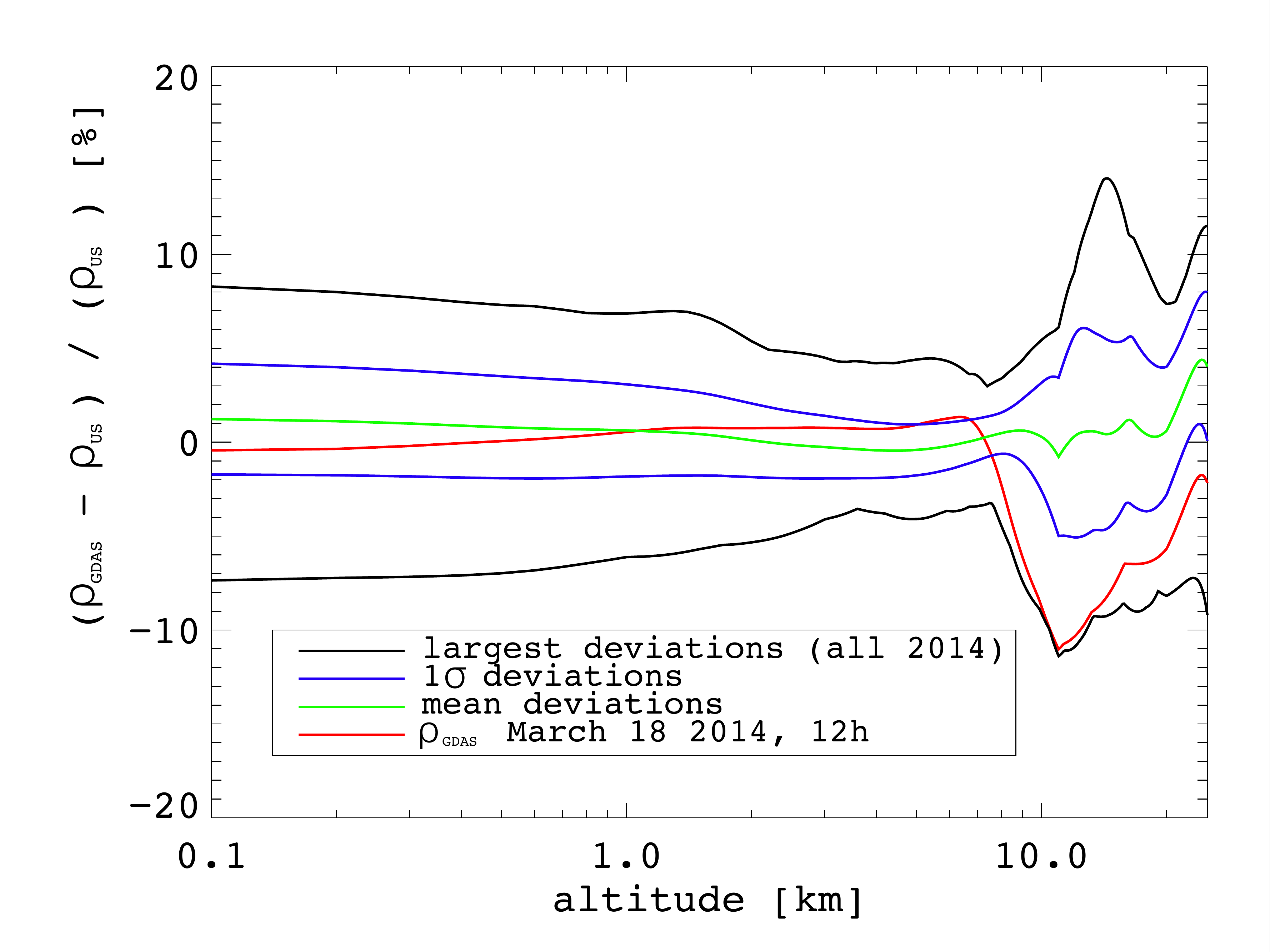}
   \caption{Top: the GDAS profile on March 18, 2014 at noon is shown in red as function of altitude along with the US Standard model in dashed blue. Bottom: the extrema of the differences between the US Standard model air density profile and all the GDAS profiles along the year 2014 are shown in black as a function of altitude. The blue lines account for the standard deviations along the year 2014 and the green one is the mean difference. In red: the GDAS profile on March 18, 2014 at noon.}
\label{fig:drho}
  \end{center}
\end{figure}

The extrema and standard deviation of the relative differences in the air density between all profiles from GDAS available in $2014$ and the US Standard model are shown in \fig~\ref{fig:drho} (bottom). We show the differences up to $\zmax$ (26 km) for better visibility, the differences being constant beyond this altitude (see \eqs~\ref{rho}). One can see that the relative difference in air density during year 2014 can reach $\pm 8$\% below $8$~km, up to $15\%$ in the range~$[10;20]$~km. These deviations affect the atmospheric depths and the air refractive index.

In order to accurately compute the atmospheric depth we have to use the spherical description together with the realistic estimation of the air density. This means that for a shower arriving from "infinity" (out of the atmosphere, where $\rho=0$) up to a distance $\ell$ from the observer measured along the shower axis (see \fig~\ref{fig:sketch}), the total atmospheric depth is given by:
\begin{equation}
\begin{split}
&\bullet \text{if}~z(\ell) > \zmax: \\
& \quad \quad X_{\text{slant}}(\ell) = f_B  \int_{\ell(\zmax)}^{\infty} \rho_\text{USstd}(z(\ell')) \, \mathrm{d}\ell'\\
&\bullet \text{if}~z(\ell) < \zmax: \\
& \quad \quad X_{\text{slant}}(\ell) = f_B \int_{\ell(\zmax)}^{\infty} \rho_\text{USstd}(z(\ell')) \, \mathrm{d}\ell'+\int_{\ell}^{\ell(\zmax)} \rho_\text{GDAS}(z(\ell')) \, \mathrm{d}\ell'
\end{split}
\label{zint}
\end{equation}
where $f_B$ is the scaling factor ensuring continuity between US standard and GDAS at altitude $\zmax$. Considering the deviations between the GDAS air density profiles and the US Standard model (see \fig~\ref{fig:drho}), relatively important differences are expected for the calculation of the atmospheric depth.

These differences are quantified as a function of the geometric distance $\ell$ to the observer for various zenith angles and with a spherical description. As depicted in \fig~\ref{fig:flat_curve} (left) the air density is integrated from the limit of the atmosphere ($\rho=0$) up to the geometrical distance to an observer located at $O$, at sea level and along the shower axis. The integrations are performed following \eqs~\ref{zint}.

\begin{figure}[H]
 \begin{center}
  \includegraphics[scale=0.3]{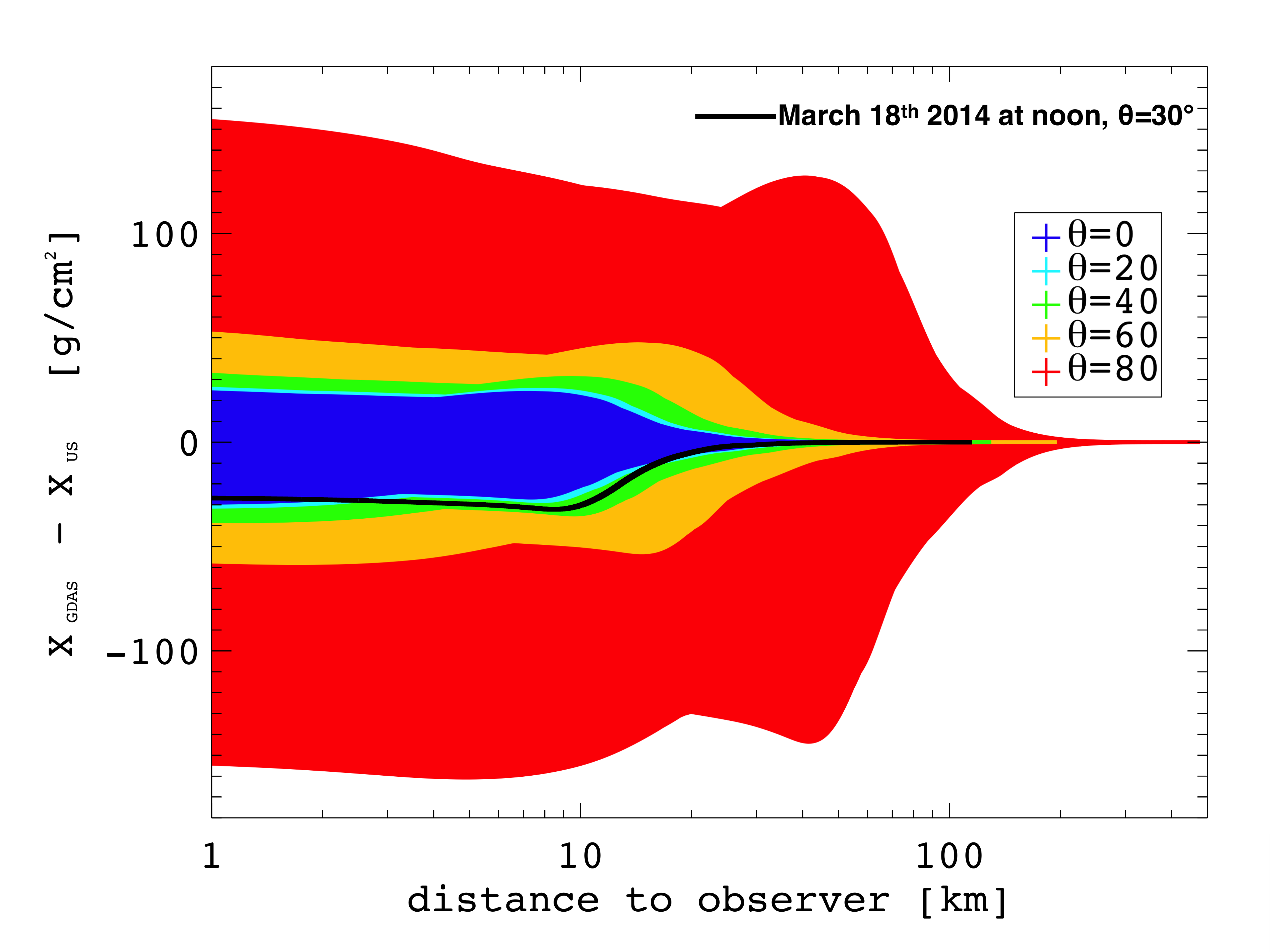}\\
    \includegraphics[scale=0.3]{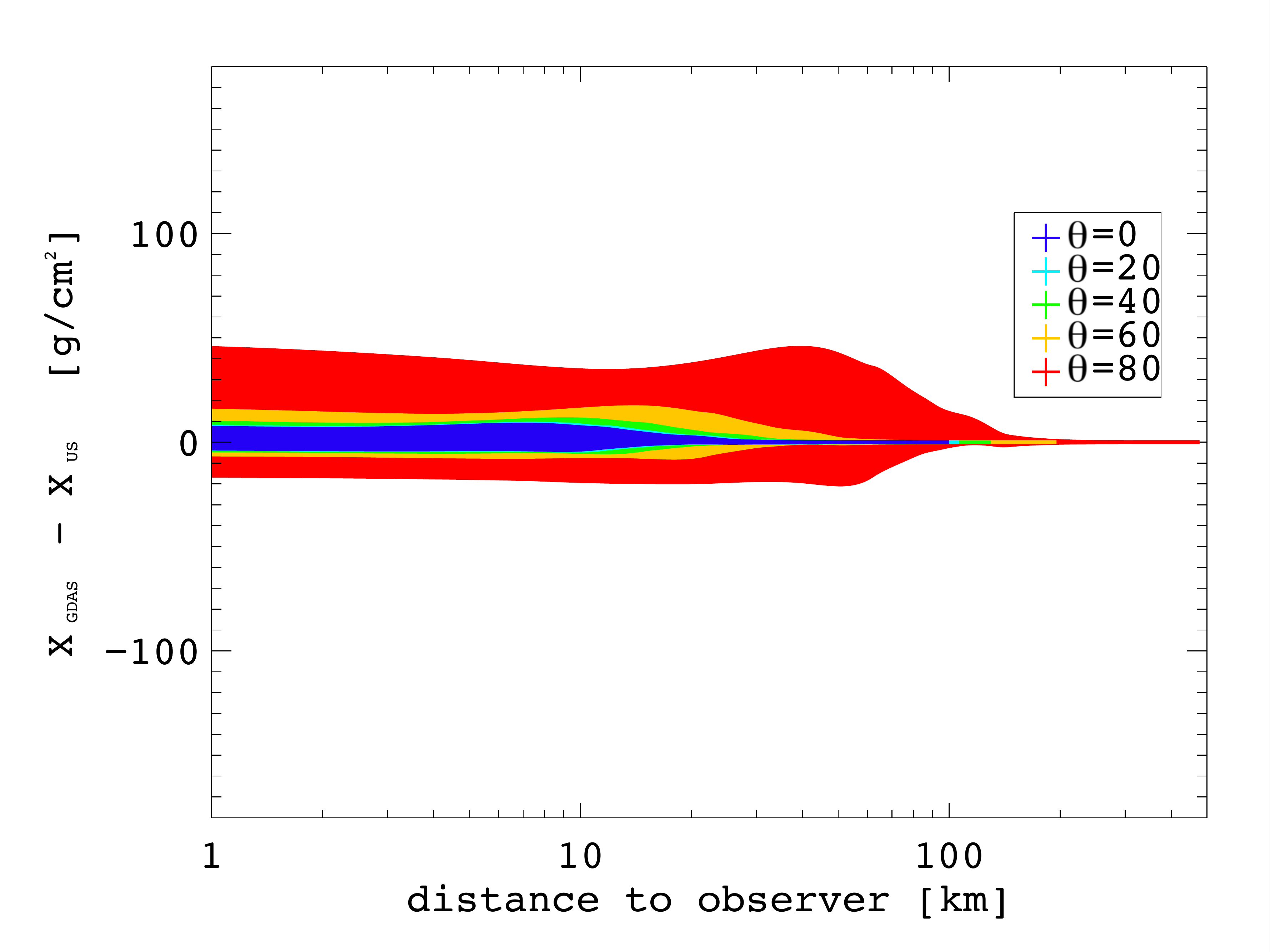}
   \caption{Top: Extrema of the atmospheric depth differences between the US Standard model and all GDAS profiles along the year~$2014$ at the location of the CODALEMA experiment as a function of source-to-observer distance $\ell$ and for various zenith angles. Bottom: The corresponding standard deviations.}
\label{fig:dx}
  \end{center}
\end{figure}

The maximum differences in the obtained crossed atmospheric depths using the GDAS profiles (up to $\zmax$ and the corrected US Standard profile beyond $\zmax$) and the US Standard model are shown in \fig~\ref{fig:dx} (top) for different zenith angles. The standard deviations of these differences are displayed in \fig~\ref{fig:dx} (bottom). 

We see that for a vertical shower the difference can be as high as $30~\gcm$ with a standard deviation of $10~\gcm$. It means that if one wants to reconstruct its \xmax\ (assuming a maximum emission at an altitude of $6$~km), using the US Standard model can induce systematic variations in atmospheric depths as high as $30~\gcm$. This is the most favorable case as these deviations are much larger for higher zenith angles and can reach $150~\gcm$ at $80^\circ$.
The aim of the radio method is to provide an accurate \xmax\ estimation and such systematic deviations must be corrected for. The use of a completely coherent description of both the atmospheric density and air refractive index is now mandatory for such analysis.

\subsection{Air refractive index}
\label{subsec:index}
The air index $\eta$ at the emission point is necessary to compute the amplitude and arrival time of the electric field emitted by secondary particles.
It depends on the air density according to the commonly used Gladstone and Dale law: 
\begin{equation}
\eta(z(l))=1+ \kappa\,\rho(z(\ell))~\text{with}~\kappa = 0.226~\text{cm}^3/\text{g}\label{glad}
\end{equation}

We also have to compute the mean air refractive index along the line of sight (between the position of the emission point and the observer's location).
This is needed to estimate the arrival time of the electric field at the antenna. This mean value is given by integrating on the line of sight with total length~$\ell$:
\begin{equation*}
<\eta(z(\ell))>=1+ \frac{\kappa}{\ell} \int_{0}^{\ell} \rho(z(\ell')) \, \mathrm{d}\ell'
\end{equation*}

However, the Gladstone-Dale constant $\kappa$ depends on the characteristics of the gas and the frequency of the light propagating in the medium. The constant $\kappa$ that was used in SELFAS and in other simulation codes like CoREAS has been determined for optical wavelengths~\cite{kappa} and is not suited to our studies in the MHz range ($\lambda=7.5$~m at $40$~MHz). As described in~\cite{gerson} the refractive index for dry air is almost constant from visible to radio wavelengths. A more consistent approach must use a description that takes into account the humidity ratio of the atmosphere. The recent formula introduced in~\cite{i3e} proposes such a description:
\begin{equation}
\eta=1+10^{-6}N\quad\text{with}\quad N = \frac{77.6}{T} \left( P + 4810\, \frac{p_v}{T} \right)\quad\quad~T~\text{in K},\label{hf}
\end{equation}
where $N$ is the refractivity. This equation is parameterized for the high and very high radio frequency range (MHz to GHz) and is suitable to our studies. In this formula, if water vapor is present, its partial pressure $p_v$ becomes dominant in the calculation of the refractive index of air. GDAS data allow to calculate the air index up to $\zmax$. Beyond this altitude, data for temperature and relative humidity are not available. However the air relative humidity beyond $\zmax$ can be taken as null: the highest clouds very rarely reach $24$~km of altitude (usually no clouds are observed above $12$~km). Thus \eq~\ref{egp} can be simplified for $z>\zmax$:
\begin{equation*}
p_v = 0,~P = p_d,~T = \frac{P_d M_d}{R\rho}\quad\text{so that}\quad N = 77.6\,\frac{R\,\rho}{M_d}\quad\text{with}\quad \rho = f_B \,\rho_\text{US}.
\end{equation*}

\begin{figure}[H]
 \begin{center}
 \includegraphics[scale=0.25]{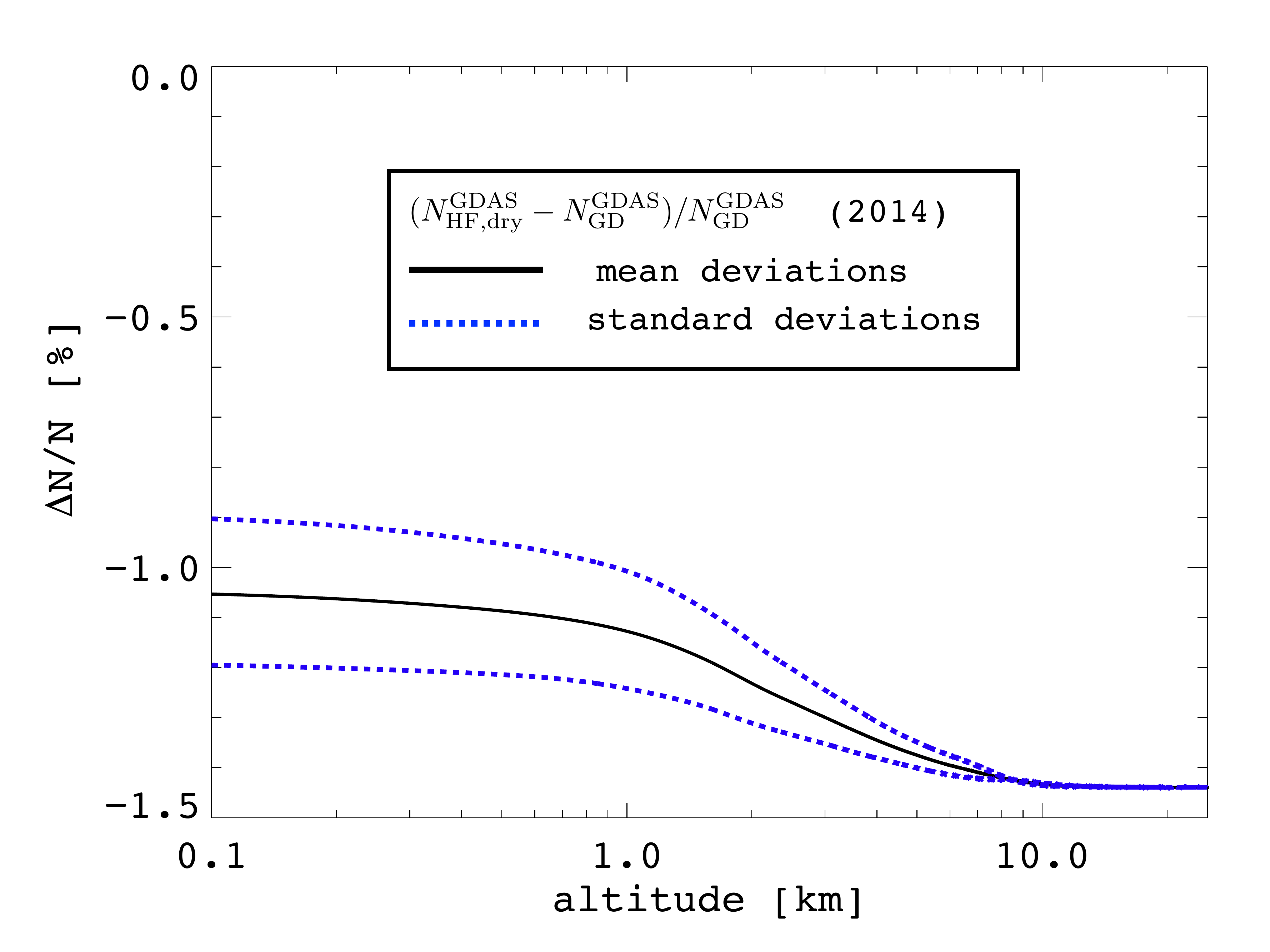}
   \caption{Relative difference of the refractivity up to $26$~km of altitude for the cases $\ngdasgd$ and $\ngdashfdry$. The black line corresponds to the mean values along the year 2014 and the blue dashed lines indicate the standard deviation of the relative difference.}
\label{fig:indair2}
  \end{center}
\end{figure}

Thus we can calculate the refractivity from ground level up to $100$~km using both the GDAS data in $[0;\zmax]$~km and the corrected mean US Standard values in $[\zmax;100]$~km. Using the Gladstone-Dale law as written in \eq~\ref{glad} is correct if we consider that the atmosphere is dry ($R_h=0$), as shown in \fig~\ref{fig:indair2}, where the relative differences between the cases $\ngdasgd$ and $\ngdashfdry$ are smaller than 1.5\%.

\begin{figure}[H]
 \begin{center}
 \includegraphics[scale=0.3]{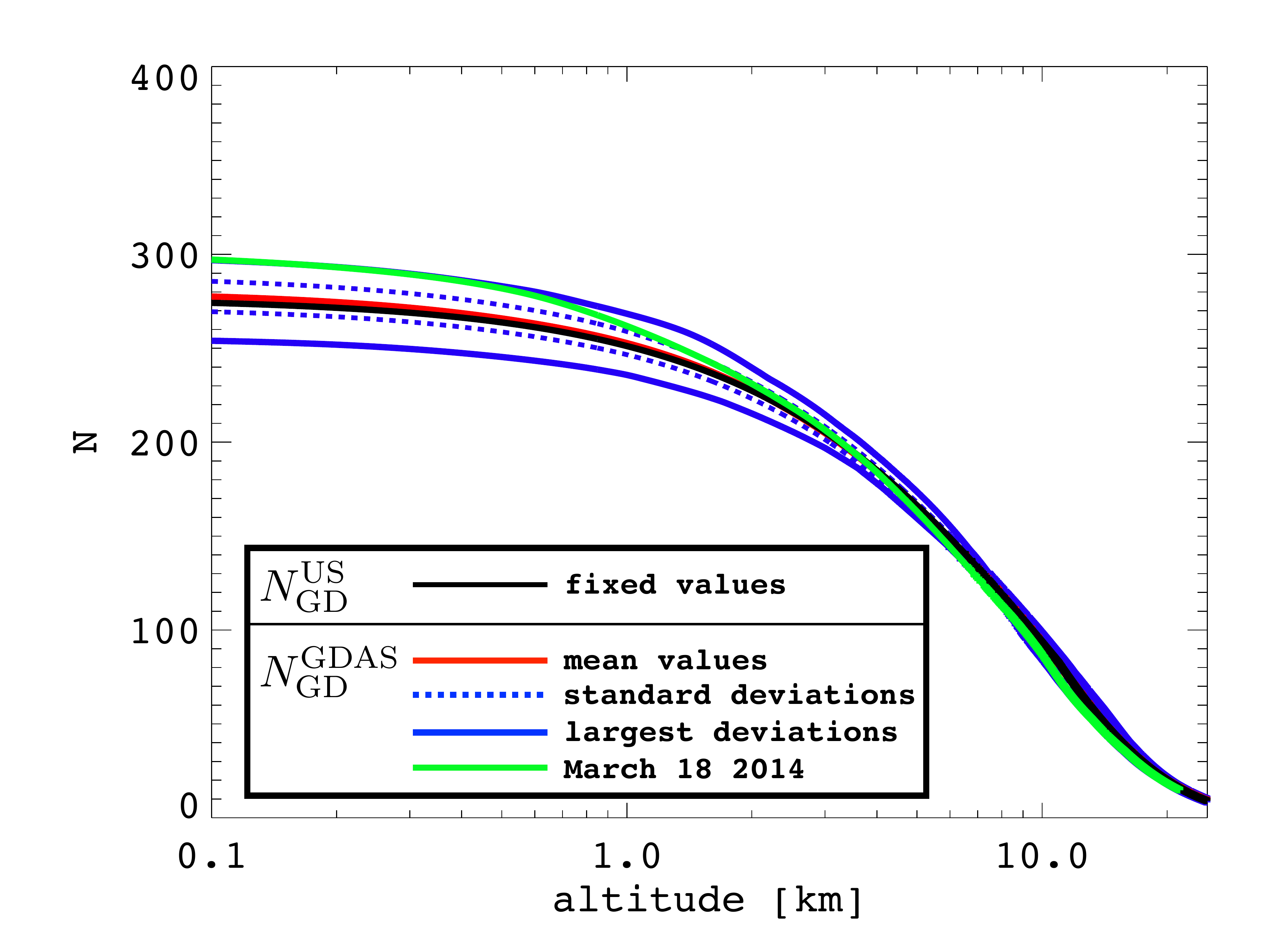}
  \includegraphics[scale=0.3]{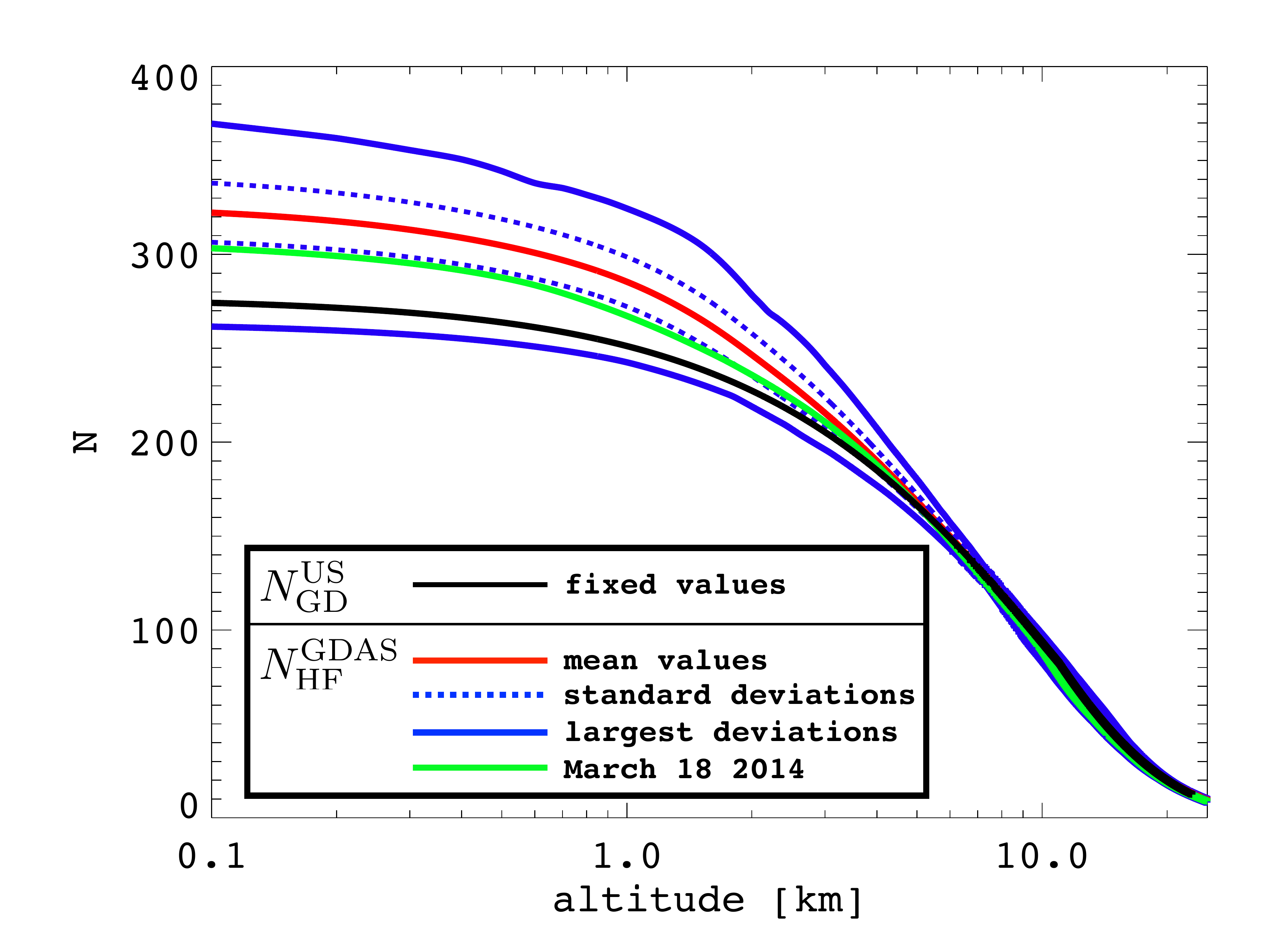}
   \caption{Refractivity up to $26$~km of altitude. The black line corresponds to the case~$\nusgd$. The blue solid lines correspond to the maximum deviations along year 2014 for the cases $\ngdasgd$ and $\ngdashf$ in the top and bottom figures, respectively. The dashed blue lines correspond to the respective standard deviations. The green lines are the values for March 18, 2014 at noon.}
\label{fig:indair}
  \end{center}
\end{figure}

\begin{figure}[H]
 \begin{center}
 \includegraphics[scale=0.3]{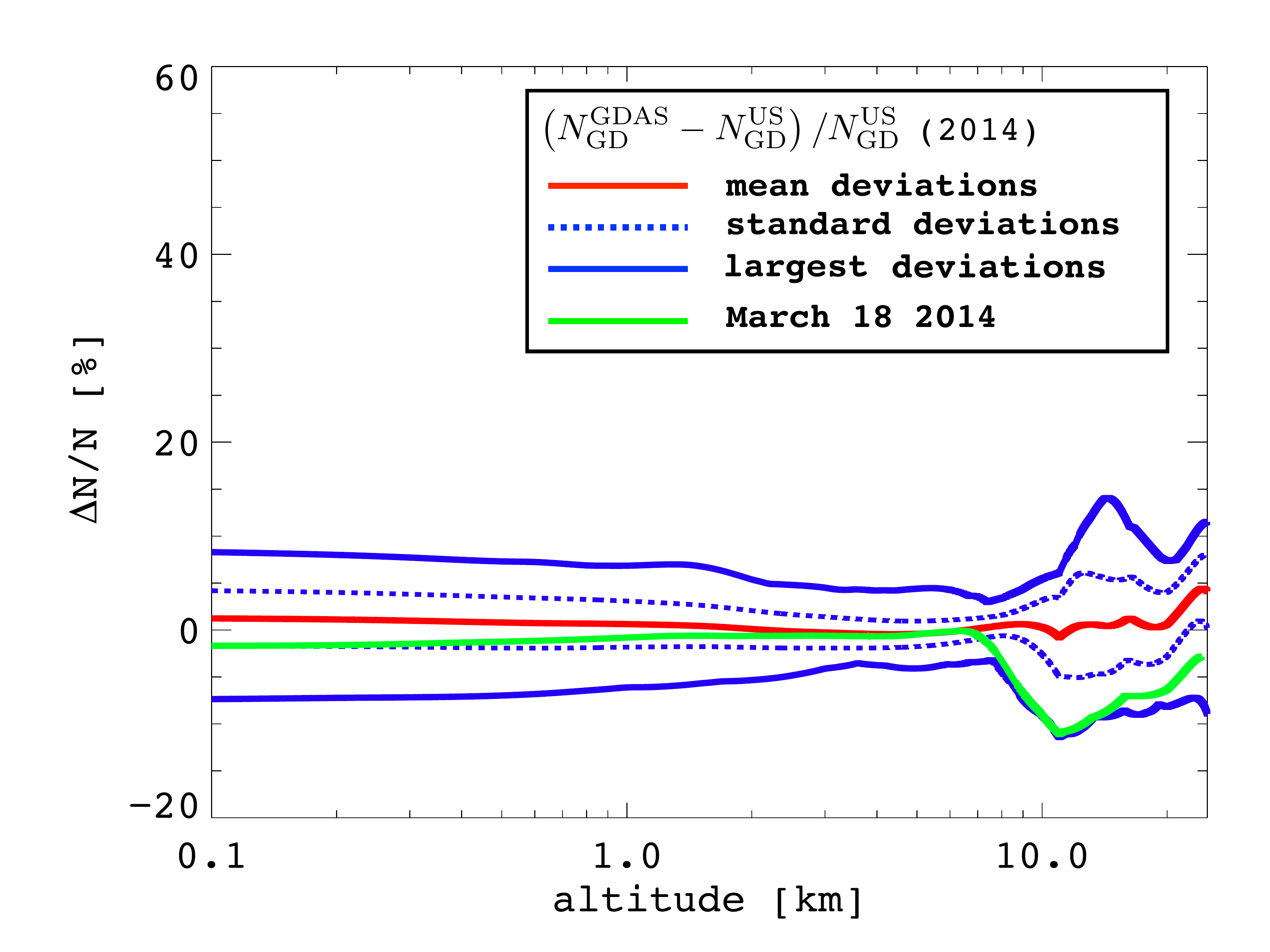}
  \includegraphics[scale=0.3]{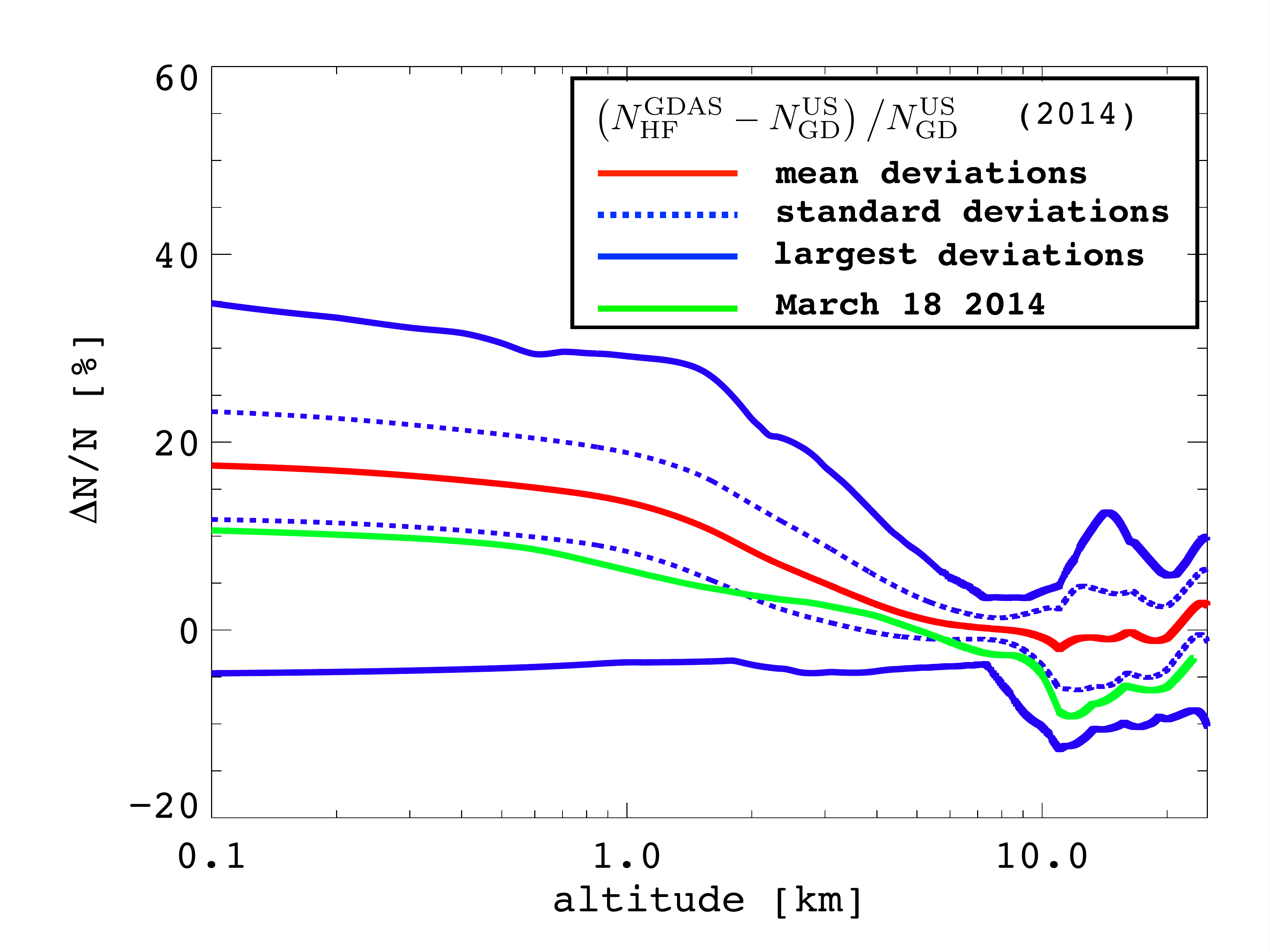}
   \caption{Relative difference in refractivity as a function of the altitude, with respect to the case $\nusgd$, for $\ngdasgd$ (top) and $\ngdashf$ (bottom). The red line corresponds to the mean values for the year 2014, the plain blue lines correspond to the maximum deviations and the dashed blue lines correspond to the standard deviations. The green lines correspond to the $\ngdasgd$ and $\ngdashf$ cases (top and bottom respectively) for March, 18, 2014 at noon.}
\label{fig:dn}
  \end{center}
\end{figure}

In order to check the importance of the choice of the air refractivity model, we consider several cases:
\begin{itemize}
\item the less refined and historic case, $\nusgd$: Gladstone-Dale law (\eq~\ref{glad}) with $\rho_\text{US}$, used in most of the simulation codes;
\item a bad case, $\ngdasgd$: Gladstone-Dale law with $\rho_\text{GDAS}$;
\item the best case, $\ngdashf$: high frequency law with water vapor $(P,T,R_h)_\text{GDAS}$.
\end{itemize}
In \fig~\ref{fig:indair} we compare the three cases $\nusgd$, $\ngdasgd$ and $\ngdashf$. The relative differences with respect to $\nusgd$ for the $\ngdasgd$ and  $\ngdashf$ cases are presented in \fig~\ref{fig:dn}. 

For the GDAS cases, we used the average GDAS values of the year~$2014$ for $P$, $T$, $R_h$ as a function of the altitude.
We observe that taking into account the mean water vapor fraction (see~\fig~\ref{fig:indair} red line for $\ngdashf$) changes significantly the value of the refractivity. This is true in the region where the water vapor is not negligible, i.e. below $\sim 10$~km. 

\begin{table}
\begin{center}
\resizebox{\textwidth}{!}{
\begin{tabular}{|l|c|c|c|c|}
    \hline
     altitude (km)& $(\ngdasgd-\nusgd)/\nusgd$  & $(\ngdashf-\nusgd)/\nusgd$ & $(\ngdashf-\ngdasgd)/\ngdasgd$ & $\left (\eta^\text{\tiny GDAS}_\text{\tiny HF}-\eta^\text{\tiny US}_\text{\tiny GD} \right )/\eta^\text{\tiny US}_\text{\tiny GD}$   \\
    \hline
    $0$    &$(1.3 \pm 4.35) \% $& $(18.1 \pm 24.0) \%$& $(16.5 \pm 5.0) \%$ & $(5.0 \pm 6.6) \times 10^{-5}$\\
    \hline
     $2.5$   &$ (-0.1 \pm 1.6) \% $&$ (6.5 \pm 11.0) \% $&$ (6.6 \pm 4.3) \%$ & $(1.4 \pm 2.4) \times 10^{-5}$\\
    \hline
     $5$   & $(-0.4 \pm 1.0)\% $&$( 1.4 \pm 3.5) \%$ &$ (1.76 \pm 2.2)  \% $ & $(2.3 \pm 5.8) \times 10^{-6}$\\
    \hline
     $7.5$ &$ (0.3 \pm 1.4) \%$ &$( 0.2 \pm 1.3) \% $ &$ (-0.17 \pm 1.2) \%$ &  $(1.2 \pm 16.7) \times 10^{-7}$\\
    \hline
     $10$   & $(0.3 \pm 3.2) \% $& $(-0.8 \pm 2.2) \%$&$ (-1.1 \pm 3.0) \%$ & $(-7.2 \pm 20.3)  \times 10^{-7}$\\
    \hline
     $12.5$   &$(0.6 \pm 6.1) \% $&$( -0.8 \pm 4.7) \%$&$ (1.35  \pm 5.6)\%$ & $(-6.0 \pm 29.8) \times 10^{-7}$\\
    \hline
         $15$   &$(0.6 \pm 5.3) \% $&$( -0.8 \pm 3.9) \%$ &$ (-1.4  \pm 4.8) \%$& $(-3.6 \pm 16.7) \times 10^{-7}$\\
    \hline
         $17.5$   &$(0.6 \pm 4.8) \% $&$(-0.8 \pm 3.3) \%$ &$ (-1.4  \pm 4.2)\%$& $(-2.4 \pm 9.5) \times 10^{-7}$\\
    \hline
         $20$   &$(0.6 \pm 4.0) \% $&$(-0.8 \pm 2.5) \%$ &$ (-1.4 \pm 3.4) \%$& $(-2.4 \pm 4.8) \times 10^{-7}$\\
    \hline
\end{tabular}}
\end{center}
\caption{Relative differences of the refractivity ($N$) and the refractive index ($\eta$) between several GDAS-based and US Standard-based models  for several altitudes of interest for air showers physics. For GDAS-based models all the data of the year $2014$ at Nan\c{c}ay were used to compute the mean differences along the year. The errors show the standard deviation at each altitude.}\label{tabn}
\end{table}
The maximum errors on the refractivity that can be induced when using $\nusgd$ instead of the more realistic $\ngdashf$ can be as large as $35\%$ close to the ground and around $15\%$ (see \fig~\ref{fig:dn}, bottom) at altitudes of interest for the shower development (below $\sim 20$~km). These values are computed for the year $2014$ but the orders of magnitude should be stable over the years. The main results of the comparison, between sea level and $20$~km of altitude are summarized in \tab~\ref{tabn}. In all columns, the relative difference using the mean of the GDAS data is displayed and the $\pm$ limits correspond to the standard deviations from the mean of the GDAS data.

\tab~\ref{tabn} implies that if we choose the most refined model $\ngdashf$, at each altitude there is a mean difference with $\nusgd$ of a few percent at the altitudes where the bulk of the shower particles lies. Since showers develop along a large range of altitudes, each layer has an impact on the travel times of the wave, which accumulates as the wave goes through each layer. These differences change the arrival time of the electric fields at the antenna, which modifies the coherence and will in turn modify also the amplitude of the electric field. Therefore, the refractive index model chosen will present different arrival times, coherence and amplitude, but with differences of a few percent (as shown in \tab~\ref{tabn}): we do not expect a drastic change.

\section{Influence of atmospheric conditions on the electric field}
\label{sec:influence}

In this section, all the simulated electric field distributions correspond to a shower initiated by a 1 EeV proton, with a first interaction depth of 27~$\gcm$ (\xmax\ = 727 $\gcm$), with an a arrival direction $(\theta,\phi) = (30,45)^\circ$ and with the geomagnetic field corresponding to Nan\c{c}ay. The simulated antennas are located along 16 directions around the shower core at ground level, there are 150 antennas in each direction with a spacing of 2 meters. The one dimensional LDFs are shown as a function of the relative distance to the shower axis along the late-early direction, i.e. from the South-West to the North-East line, which is the direction having the same azimuth as the incoming shower. The positive axis distances correspond to early positions relative to the shower core and the negative values correspond to late positions.

\subsection{Air density profile}
\label{subsec:efieldrho}

\begin{figure}[H]
 \begin{center}
  \includegraphics[scale=0.4]{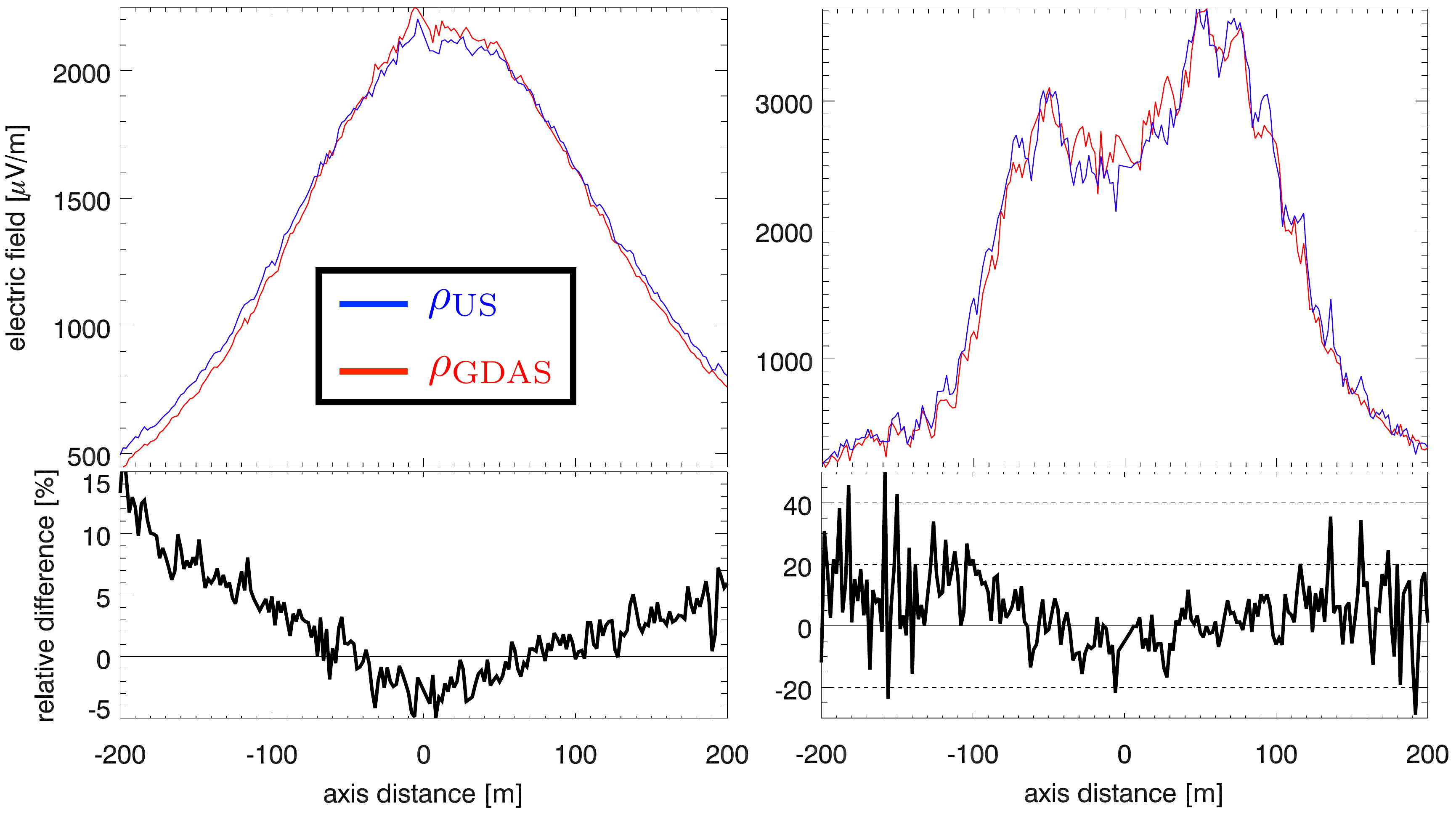}
   \caption{Top: Total electric field amplitude as a function of distance to the shower axis. The electric field is filtered in the band $[20;80]$~MHz (left) and $[120;250]$~MHz (right). We used the US Standard model (blue curve) and the atmospheric model based on the GDAS data on March 18, 2014 at noon (red curve). Bottom: the relative amplitude differences between the two air density profiles as a function of axis distance for the corresponding frequency bands.}
\label{fig:ldfdensity}
  \end{center}
\end{figure}

The influence of the atmospheric model is explicit in the total electric field amplitude, i.e. the lateral distribution function (LDF) in $[20;80]$~MHz and $[120;250]$~MHz, as shown in~\fig~\ref{fig:ldfdensity}. The LDF is calculated as the maximum of quadratic sum of the three polarizations. 

In this figure, we simulated a shower using both the US Standard model (in blue) and the GDAS realistic conditions of March 18, 2014 at noon (in red). We observe that the relative difference between the two LDFs varies as a function of the distance to the shower axis. This implies that the maximum emission occurs at lower altitude when using the GDAS profile. This leads to a systematic error when trying to reconstruct the \xmax\ using a model with a constant atmosphere.

\subsection{Air refractive index}
\label{subsec:efieldN}

In order to quantify the differences in the electric fields induced by a change in the refractive index,
the time traces for the three polarizations have been simulated with SELFAS for March 18, 2014 at noon, for different antenna axis distances.

\begin{figure}[H]
 \begin{center}
  \includegraphics[scale=0.33]{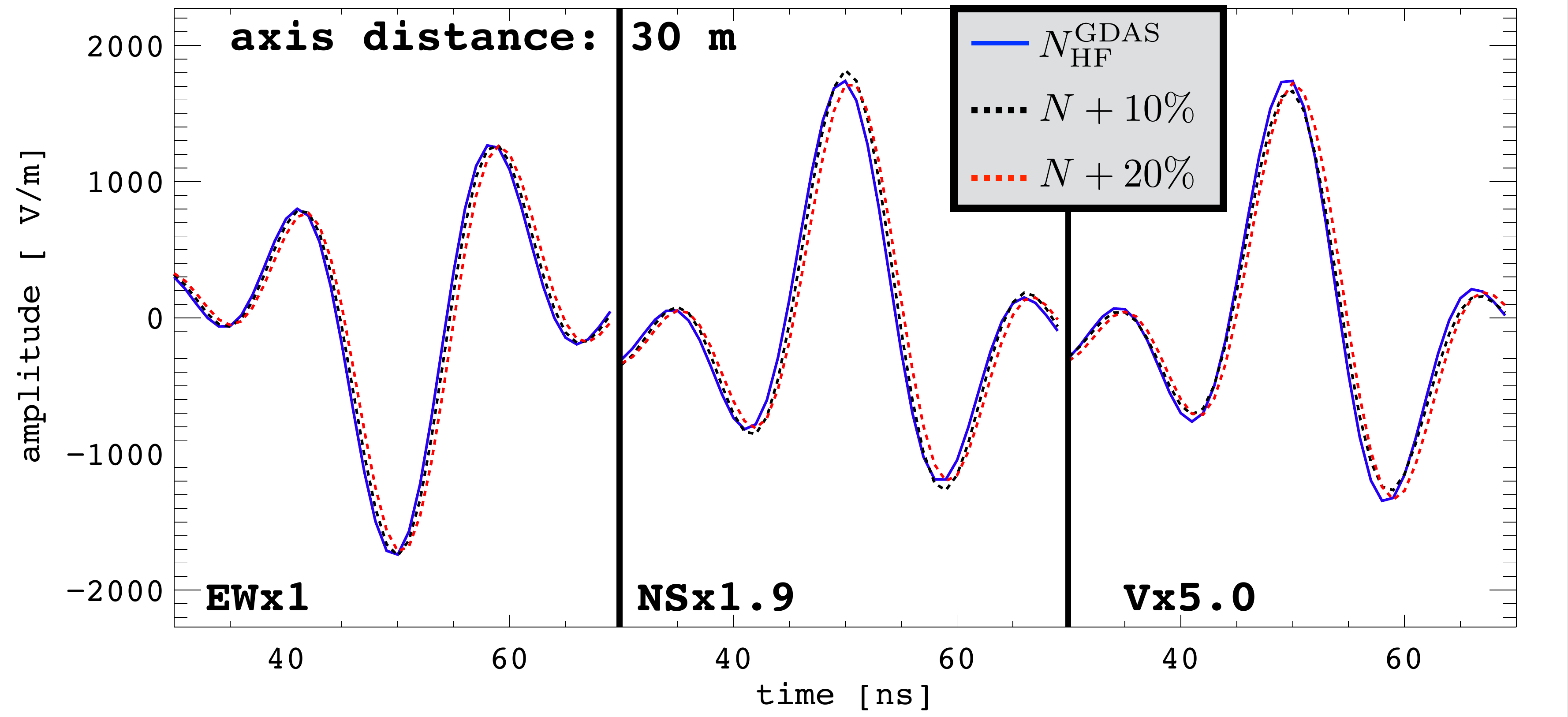}
    \includegraphics[scale=0.33]{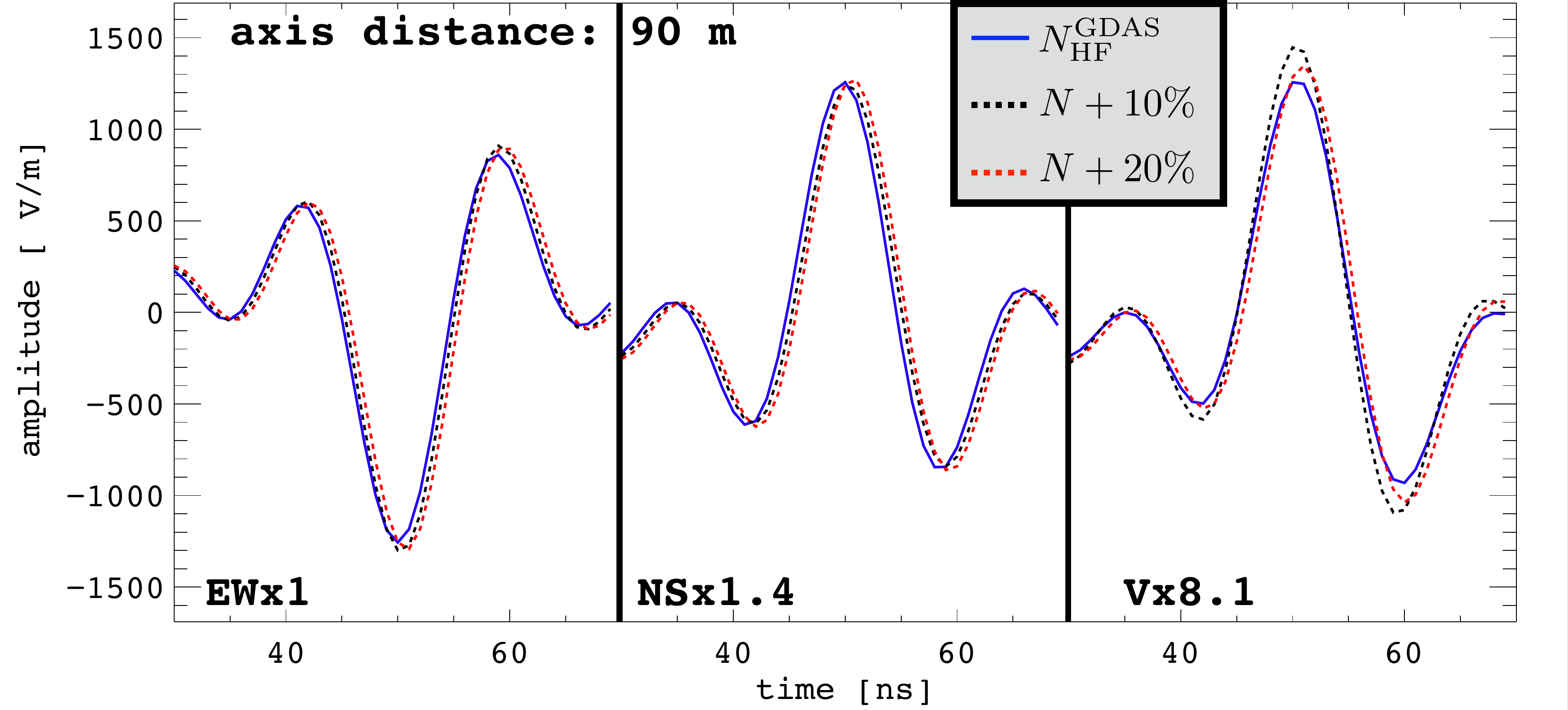}
      \includegraphics[scale=0.33]{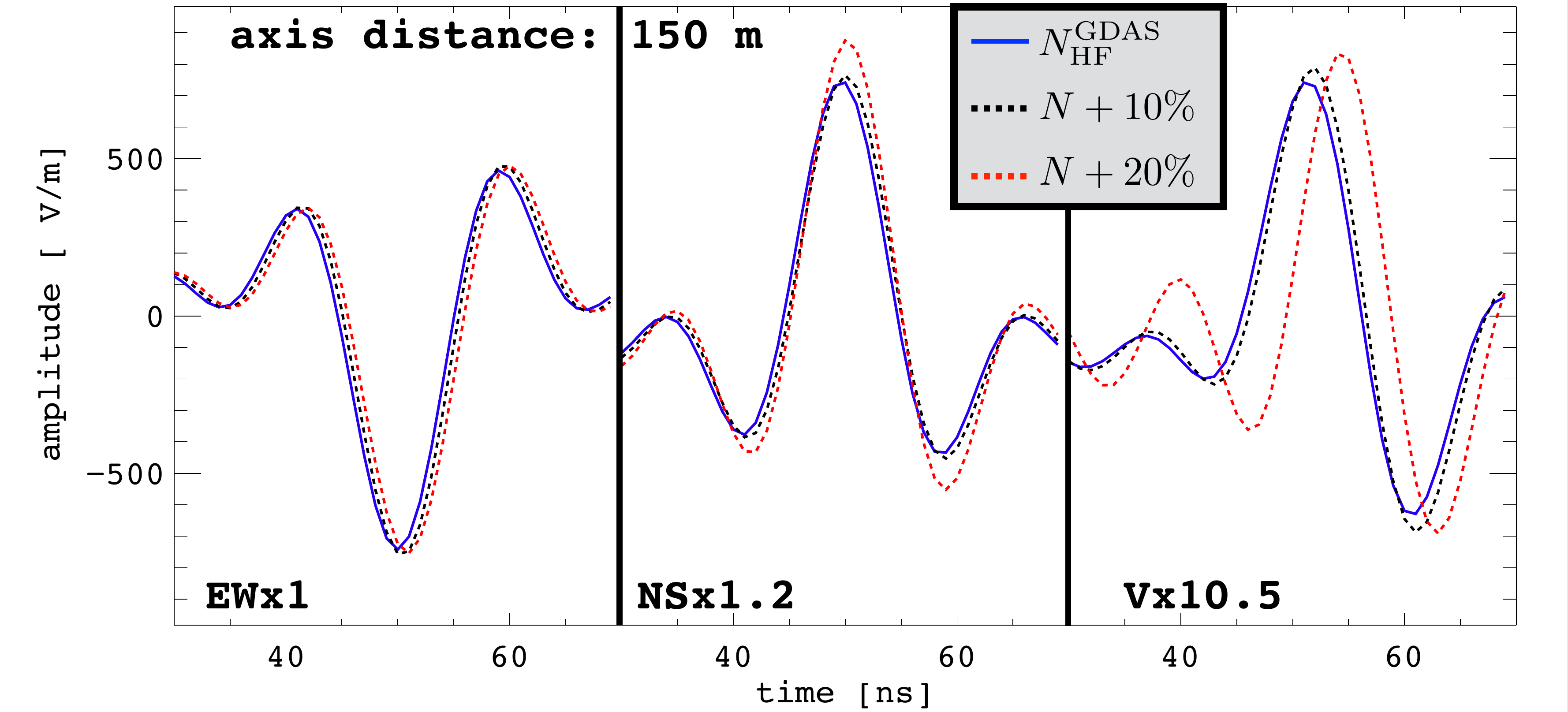}
   \caption{Time traces of the electric field in the late (at the south-west of the core) direction simulated with SELFAS using the refractive index model $\ngdashf$ (in blue), $\ngdashf+10\%$ (in black) and $\ngdashf+20\%$ (in red), with the same air density profile, for different distances to the shower axis in the shower front reference frame. The electric field is filtered in the band $[20;80]$~MHz for the three polarizations, indicated at the bottom of each plot together with the scale factor for better visibility.}
\label{fig:trace}
  \end{center}
\end{figure}

 We have used three different air indexes, namely, the GDAS HF model $\ngdashf$, the same model with a $10\%$ increase in the refractive index ($\ngdashf+10\%$) and with a $20\%$ increase ($\ngdashf+20\%$), while the air density profiles have been kept identical and have been calculated using \eq~\ref{zint}. The showers are completely identical, the same seed has been used in SELFAS to draw the energy, position and speed of the secondary particles. The results, filtered in the $[20;80]$~MHz band, are shown in~\fig~\ref{fig:trace}. The amplitudes have been multiplied by a factor indicated at the bottom of each plot for better visibility.

\begin{figure}[H]
 \begin{center}
  \includegraphics[scale=0.33]{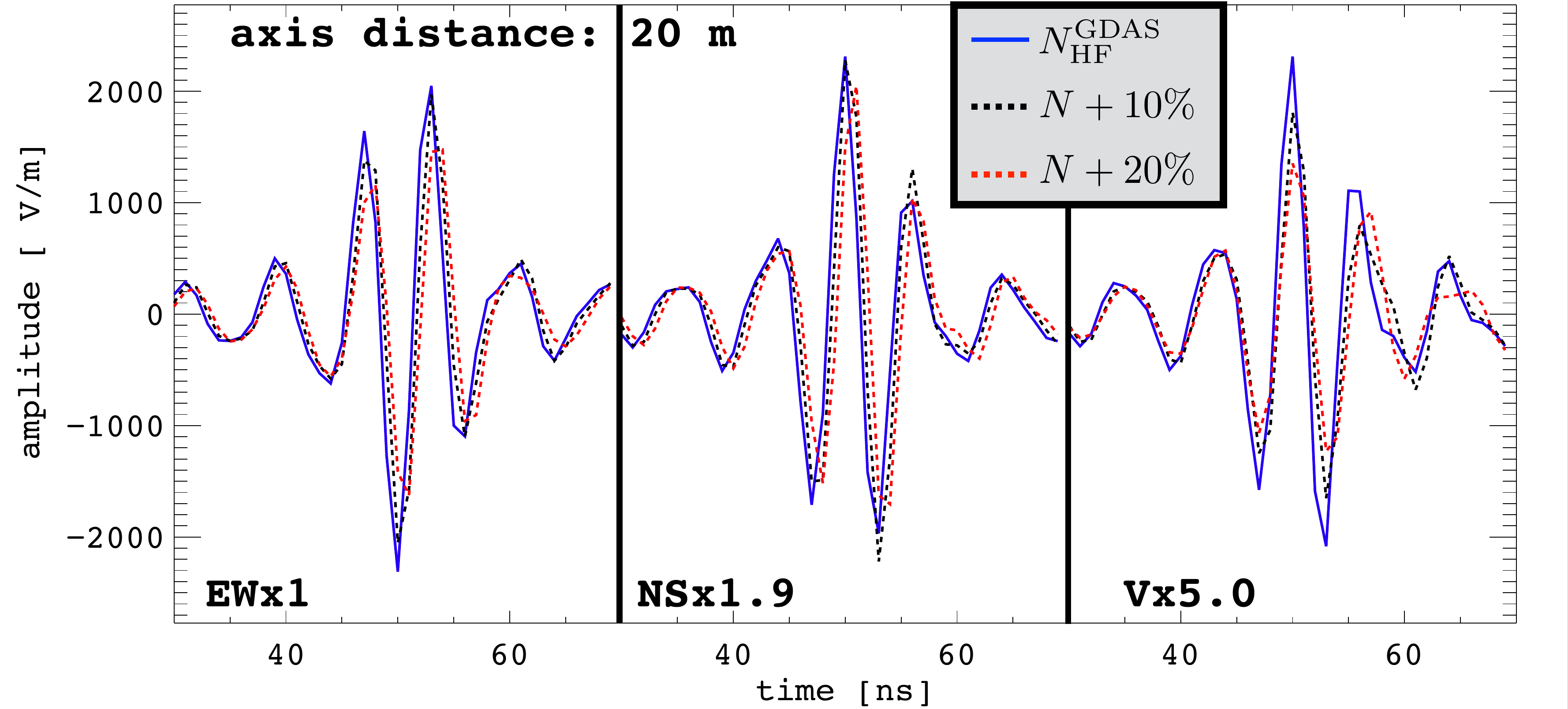}
    \includegraphics[scale=0.33]{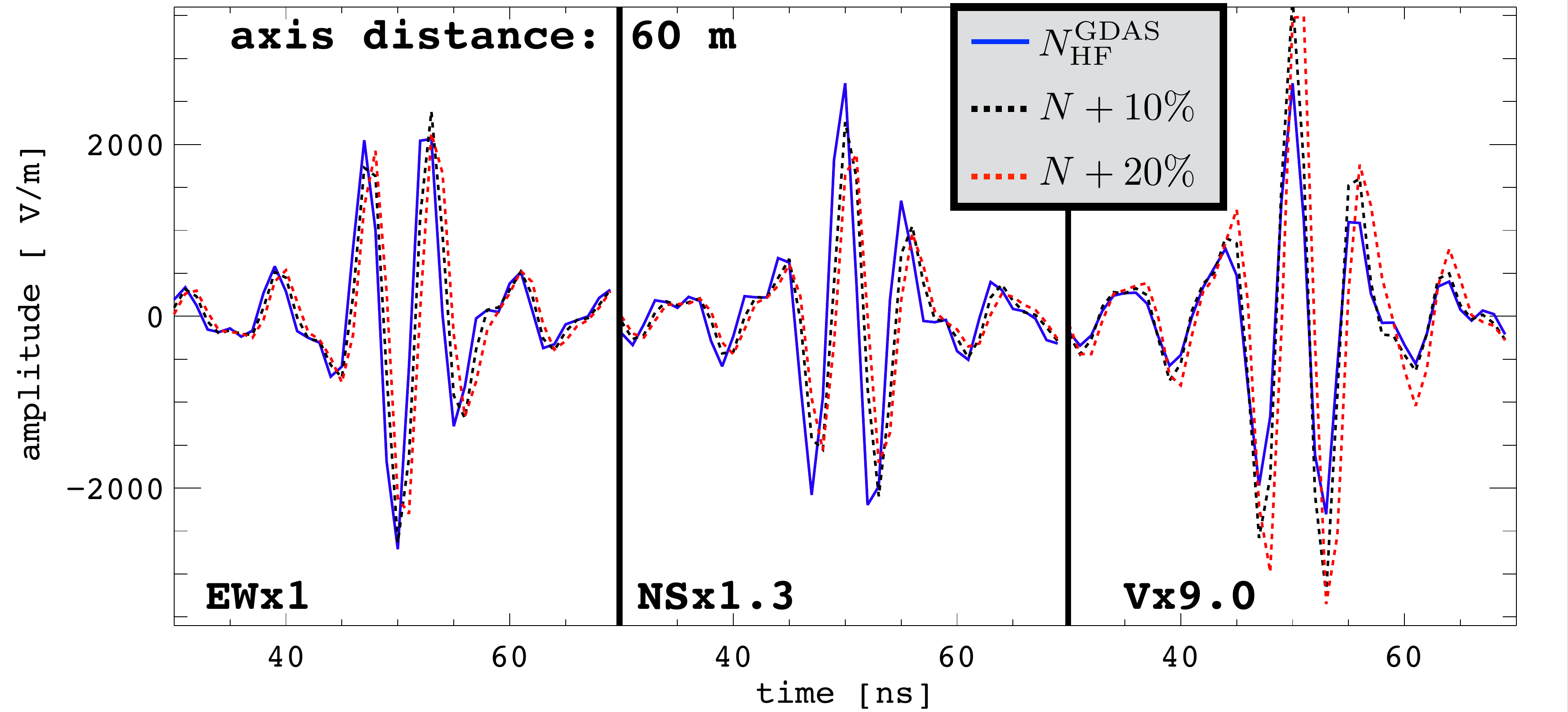}
      \includegraphics[scale=0.33]{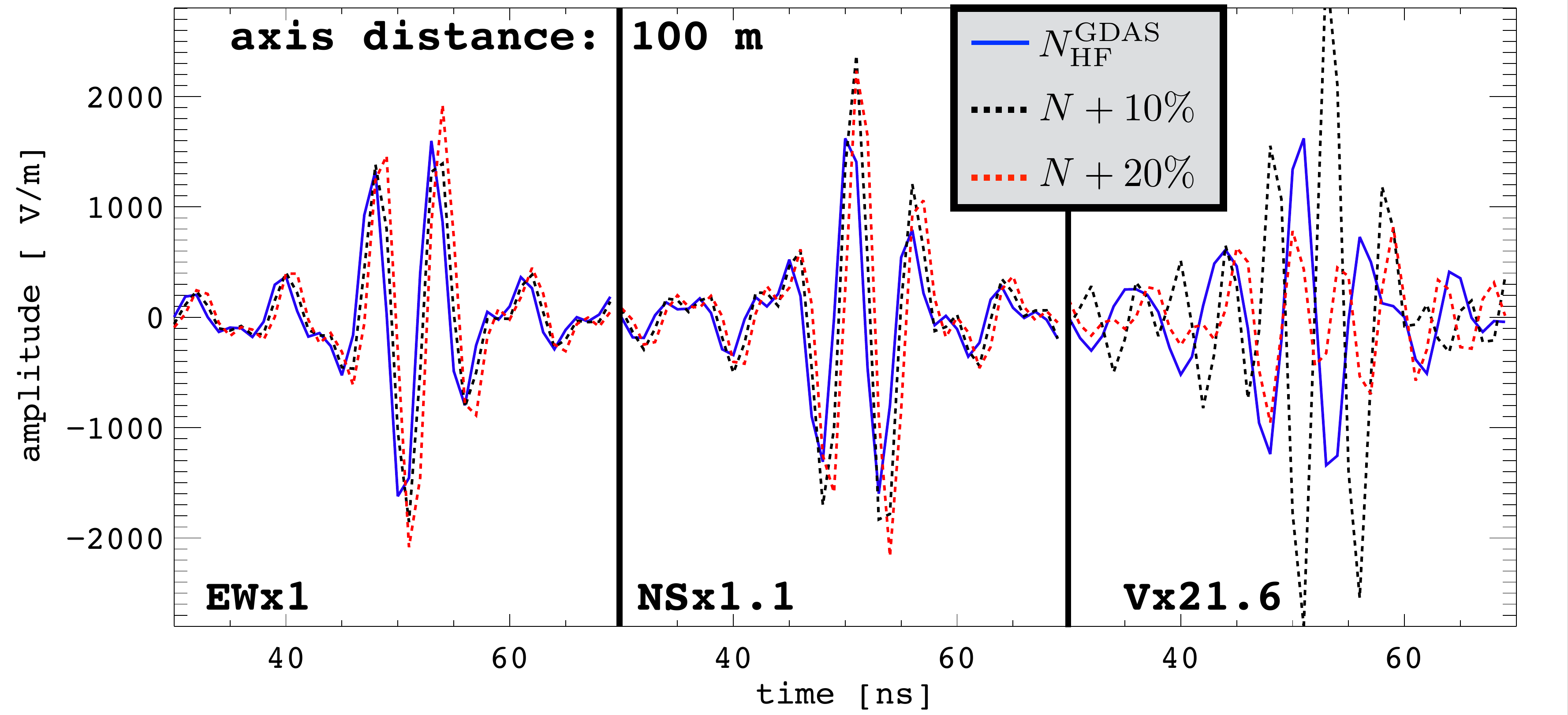}
   \caption{Same as \fig~\ref{fig:trace} in the $[120;250]$~MHz band.}
\label{fig:trace0}
  \end{center}
\end{figure}

We see that for the inspected antennas, in the $[20;80]$~MHz band, differences in the maximum amplitude are of a few percent but vary as a function of the axis distance: 0\% to +6\% between $\ngdashf$ and $\ngdashf+10\%$ and +2\% to +8\% between $\ngdashf$ and $\ngdashf+20\%$.  Differences are of a few ns in the arrival time of the maximum: +1 ns between $\ngdashf$ and $\ngdashf+20\%$, for the EW and NS polarizations and +4 ns for the vertical polarization at an axis distance of 150 m. The different amplitudes predicted by the three models imply that the footprints on the ground may be different. The variation of the relative differences of the LDFs as a function of the axis distance can affect the shower maximum reconstruction, as discussed in section~\ref{subsec:oneevent}. An absolute (overall) time shift in the pulses is generally not important, as one uses relative timings. Relative time differences between antennas may, and do in fact, occur, although they are not relevant for the present work's estimation of the shower maximum, as one uses the amplitudes only. The picture changes if we increase the observation frequency. We show in
\fig~\ref{fig:trace0} the same plot than in \fig~\ref{fig:trace}, but filtered in the
$[120;250]$~MHz band. In this case, the differences between the three refractivity profiles are
more pronounced. At these frequencies, we chose antennas closer to the axis as the electric field is no longer emitted coherently beyond 100 meters.

\begin{figure}[H]
 \begin{center}
  \includegraphics[scale=0.4]{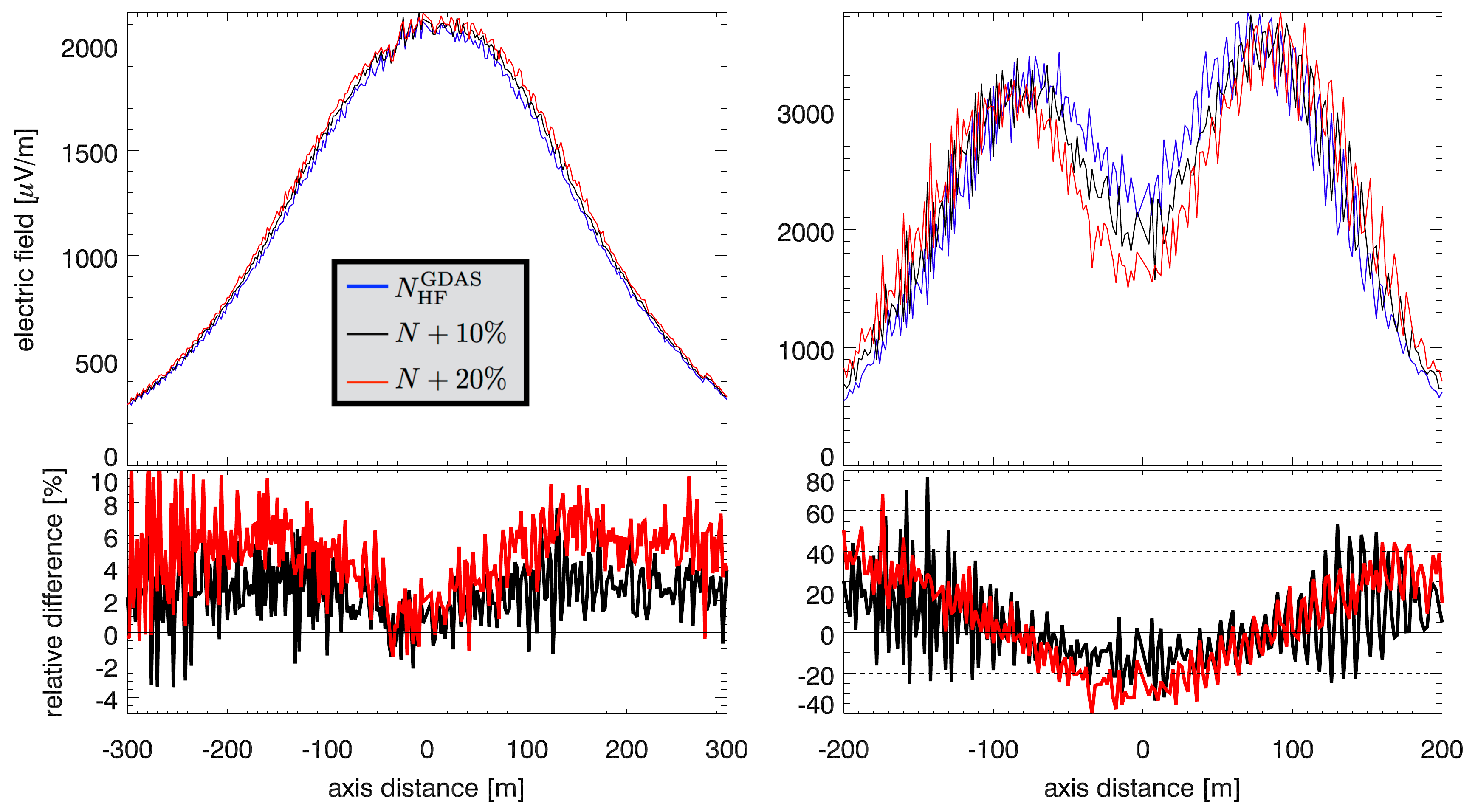}
   \caption{Top left: total electric field amplitudes simulated with the refractive index $\ngdashf$ (in blue), $\ngdashf+10\%$ (in black) and $\ngdashf+20\%$ (in red) as a function of the distance to the shower axis, using the same simulated shower. The electric field is filtered in the $[20;80]$~MHz band. Bottom left: the corresponding relative differences (with respect to the case $\ngdashf$) of the amplitude of the electric field at a maximum distance of $200$~m from the shower axis, where the emission of the electric field is coherent. Right: Same as left, but for the $[120;250]$~MHz band.}
\label{fig:ldfindex}
  \end{center}
\end{figure}

We show in \fig~\ref{fig:ldfindex} the LDF for the maximum of the electric field calculated with the three different refractive indexes. We check that, in the
$[20;80]$~MHz band (\fig~\ref{fig:ldfindex}, left), differences for an increase of even $20\%$ in the refractive index amounts
to an amplitude error of $\sim 5\%$. However, we find that for a higher frequency band ($[120;250]$~MHz, see \fig~\ref{fig:ldfindex} (right)) differences in the electric field maximum amplitude can reach up to $40\%$. Besides, the asymmetry on the LDF both sides of the shower axis is remarkable.

\begin{figure}[H]
 \begin{center}
  \includegraphics[scale=0.38]{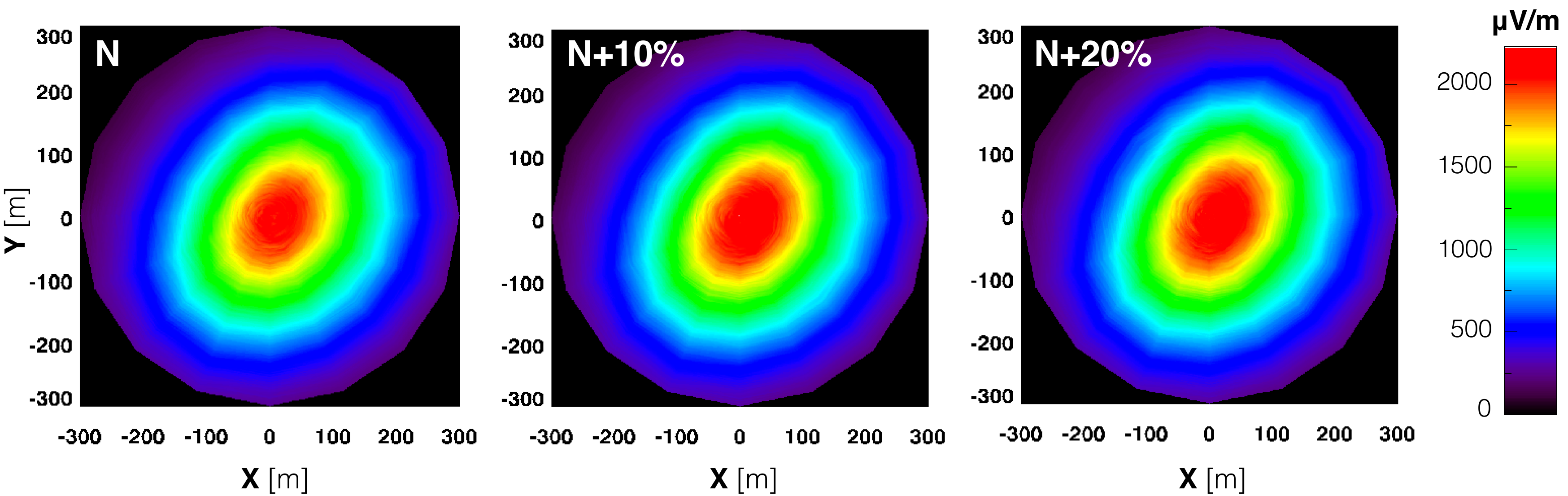}
   \caption{Two-dimensional lateral distribution functions in the ground reference frame
for $\ngdashf$ (left), $\ngdashf+10\%$ (middle) and $\ngdashf+20\%$ (right) filtered in the band $[20;80]$~MHz for the same shower of \fig~\ref{fig:ldfindex}. The $X$ coordinate represents the easting of the observer and the 
   $Y$ coordinate represents the northing of the observer. All the antennas have been placed at
   ground level.}
\label{fig:compN2080}
  \end{center}
\end{figure}

Plotting the two-dimensional distribution of the maximum
of the electric field, as we do in \fig~\ref{fig:compN2080}, confirms that the distribution for
differences of $10\%$ and $20\%$ in the refractive index amounts to a feeble difference
that is almost imperceptible with the naked eye in the $[20;80]$~MHz band. The Cherenkov
ring in this frequency band is not expected to behave as a typical Cherenkov ring, since
even for low frequencies the lateral extension of the shower is important for the calculation
of the arrival times and field coherence \cite{zhairesicrc}. That is why a $20\%$ difference
in the refractive index does not appreciably change the Cherenkov ring at low frequency.

\begin{figure}[H]
 \begin{center}
  \includegraphics[scale=0.38]{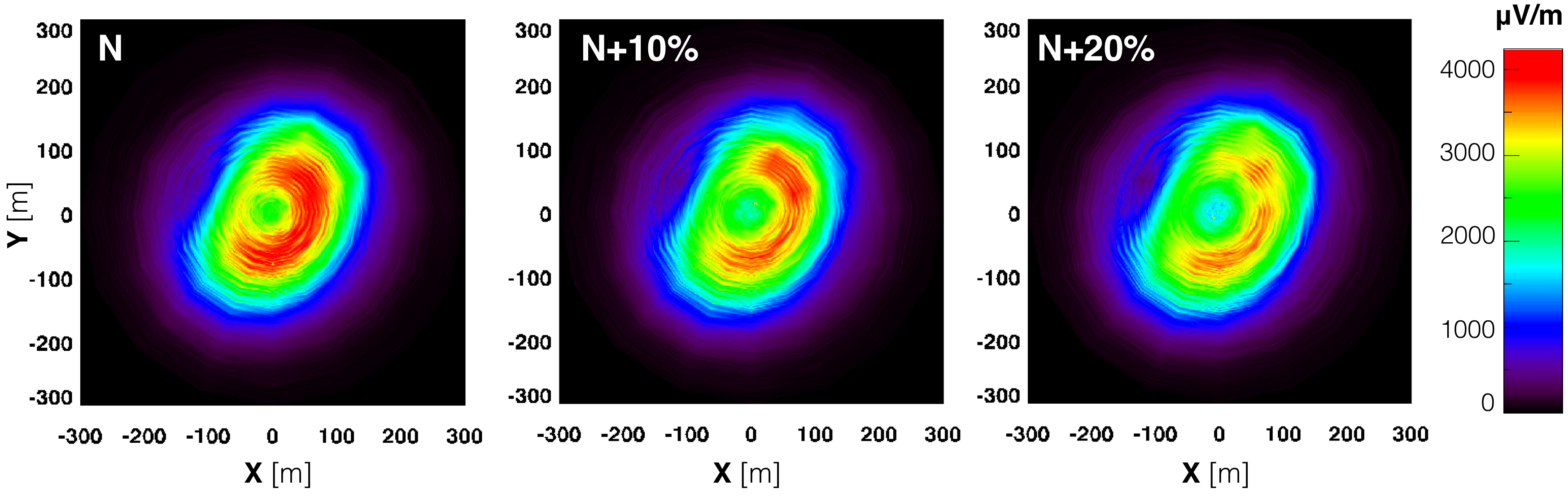}
   \caption{Same as \fig~\ref{fig:compN2080} in the $[120;250]$~MHz band.}
\label{fig:compN}
  \end{center}
\end{figure}

The two-dimensional lateral distribution functions in the $[120;250]$~MHz band are shown in \fig~\ref{fig:compN} where not only 
the Cherenkov ring is evident (as expected at these frequencies), but also it moves
when we change the refractive index. 
The ground distribution of the electric field presents an elliptical asymmetry created
by the intersection of the Cherenkov cone with the ground plane.

We can affirm, then, that a change of even $20\%$ in the refractive index propagates
into a difference of a few percent in the amplitude of electric field in the $[20;80]$~MHz
band, while in the $[120;250]$~MHz band the changes become quite drastic (tens of percent). We conclude
that lower frequencies seem to be less sensitive to changes in the refractive index. It means that if one measures the electric field at high frequencies, caution must be taken with the air index model.

For the $\ngdashf$ model in the $[120;250]$~MHz band (\fig~\ref{fig:compN}, left), the Cherenkov ring is located at a distance of $72$~m from the shower core in the ground reference frame, towards the late direction;
the ring radius increases to $79$~m for $\ngdashf+10\%$ and to $89$~m for
$\ngdashf+20\%$.
These radii of the Cherenkov rings are obtained by fitting the LDF  with a 1D gaussian function in the late direction (along the long axis of the Cherenkov ellipse). 
As a sanity check, we compare the obtained values with a simple modeling of the Cherenkov radius in the late direction. The angle $\theta_C$, with respect to the shower axis, at which the Cherenkov emission occurs is given by: $\theta_C = \text{acos} (1/(\eta \beta))$.
In the ground reference frame, the Cherenkov radius in the late direction of the shower is thus given by $R_C=D_\text{max}(\cos\theta \tan(\theta+\theta_C) - \sin\theta)$,
where $\theta=30^\circ$ and $D_\text{max} \simeq 4150$~m, is the distance between \xmax\ and the shower core in our example. The refractive index $\ngdashf \simeq 200$ corresponds to the altitude of $h_{X_{\text{max}}}= 3700$~m. We must always be aware that an EAS possesses a lateral extension,
as well as a shower front thickness, so a one-dimensional model
will not suffice, in general, to calculate the footprint on the ground,
and in particular, the size of the Cherenkov ring. However, if we
fix an observation frequency large enough so that only the particles
near the shower axis contribute coherently to the electric field, we
can expect better agreement between a one-dimensional model
and the simulations. Let us take an observation frequency equal to 300~MHz,
which implies a coherent contribution of the particles at a distance of
less than~1~m from the shower axis as done in \cite{zhairesicrc}.
\begin{table}[!ht]
\begin{center}
\begin{tabular}{|c|c|c|c|}
    \hline
     air index & $\ngdashf$& $\ngdashf+10\%$ & $\ngdashf+20\%$ \\
    \hline
    SELFAS & 96 m. & 105 m & 113 m  \\
    \hline
    Analytic & 97 m & 103 m & 107 m \\
    \hline
\end{tabular}
\end{center}
\caption{Cherenkov radii (distances from the shower core to the maximum of the LDF) 
for the South-West line of the shower in \fig~\ref{fig:compN2080} at 300 MHz,
calculated with SELFAS and with the analytical expression of the Cherenkov angle,
for different values of the air refractivity.}
\label{tabcerenkocompv}
\end{table}
We show the results for the maximum of the LDF at 300 MHz in \tab~\ref{tabcerenkocompv},
that indicate that the one-dimensional model and the simulation are in 
agreement at high frequency (better than $\sim 6\%$ for this case).

\section{Influence of atmospheric conditions on the \xmax\ estimation}
\label{sec:recoxmax}

\subsection{Example with one reconstructed event}
\label{subsec:oneevent}

In this section we show the importance to consider the actual atmospheric experimental conditions to reconstruct \xmax\ from the radio signal. The method is based on the comparison of the LDF actually sampled by an array of antennas such as CODALEMA to a set of simulated LDFs. The electric field is strongly beamed towards the direction of propagation of the shower so that the overall shape of the LDF depends on \xmax. Thus each simulated LDF is induced by a shower with a particular \xmax\ and the comparison of the experimental LDF to the simulated set allows the determination of \xmax\ giving the best agreement. To illustrate this method, an event is simulated using the conditions of the CODALEMA experiment at noon on March 18, 2014 using the GDAS data: air density profile given by $\rho_\text{GDAS}$ and air refractivity profile given by $\ngdashf$. The shower is initiated by a $1$~EeV proton, its arrival direction is $(\theta,\phi) = (30^\circ, 90^\circ)$ and $X_\text{max} = 702~\gcm$ ($X_1 = 15~\gcm$). This simulated event is considered as a test event which is compared to three simulated data sets composed of showers induced by protons and iron nuclei, with the same arrival direction and energy but random \xmax:
\begin{itemize}
\item the first set uses $(\rho_\text{US},\nusgd)$
\item the second set uses $(\rho_\text{GDAS},\ngdashf)$ on March 18, 2014 at noon
\item the third set uses ($\rho_\text{US},\ngdashf)$ on March 18, 2014 at noon.
\end{itemize}
These choices allow to check independently the influence of air density and air index profiles.

\begin{figure}[!ht]
 \begin{center}
 \includegraphics[width=9cm,height=5.2cm]{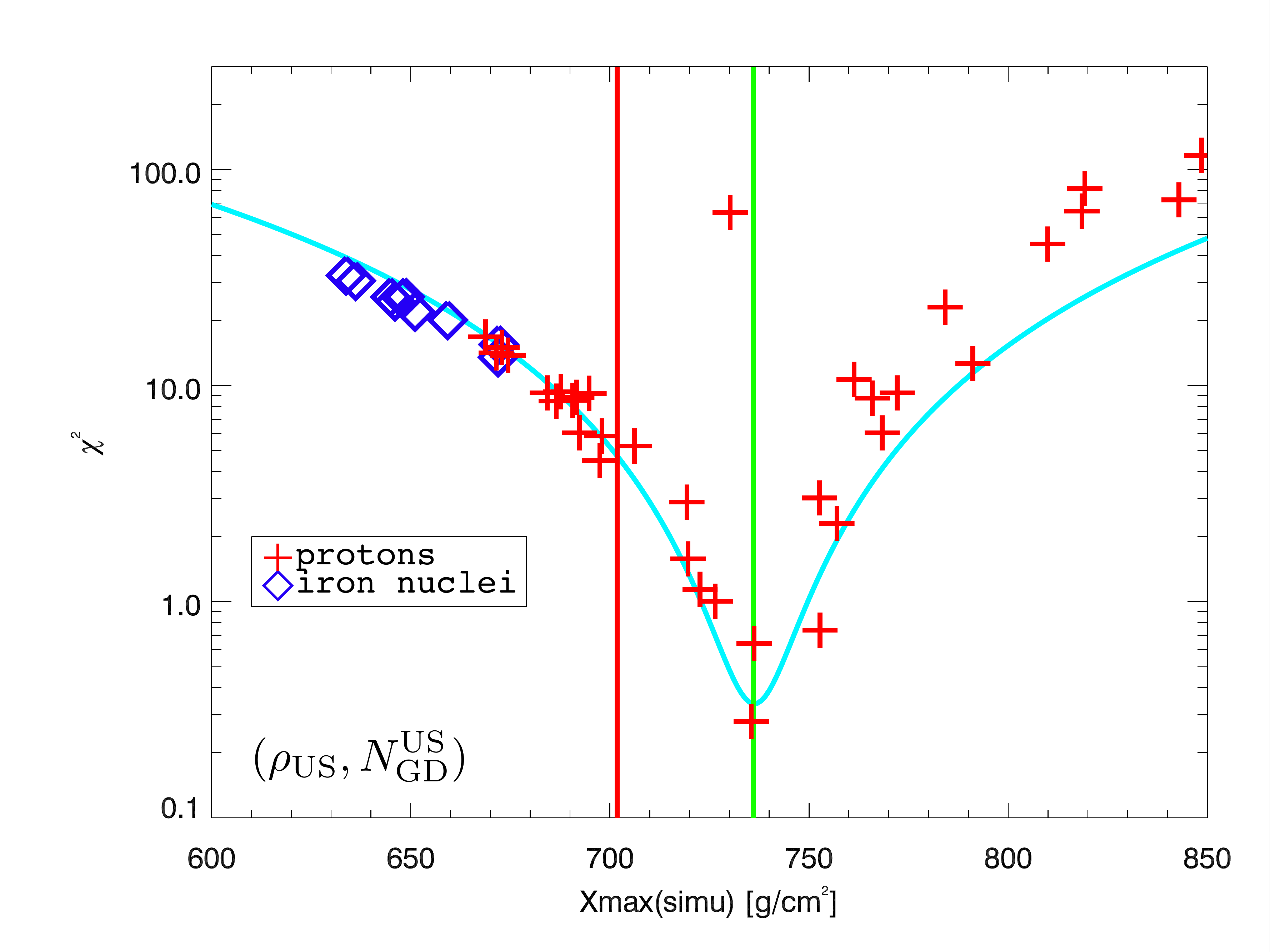}
  \includegraphics[width=9cm,height=5.2cm]{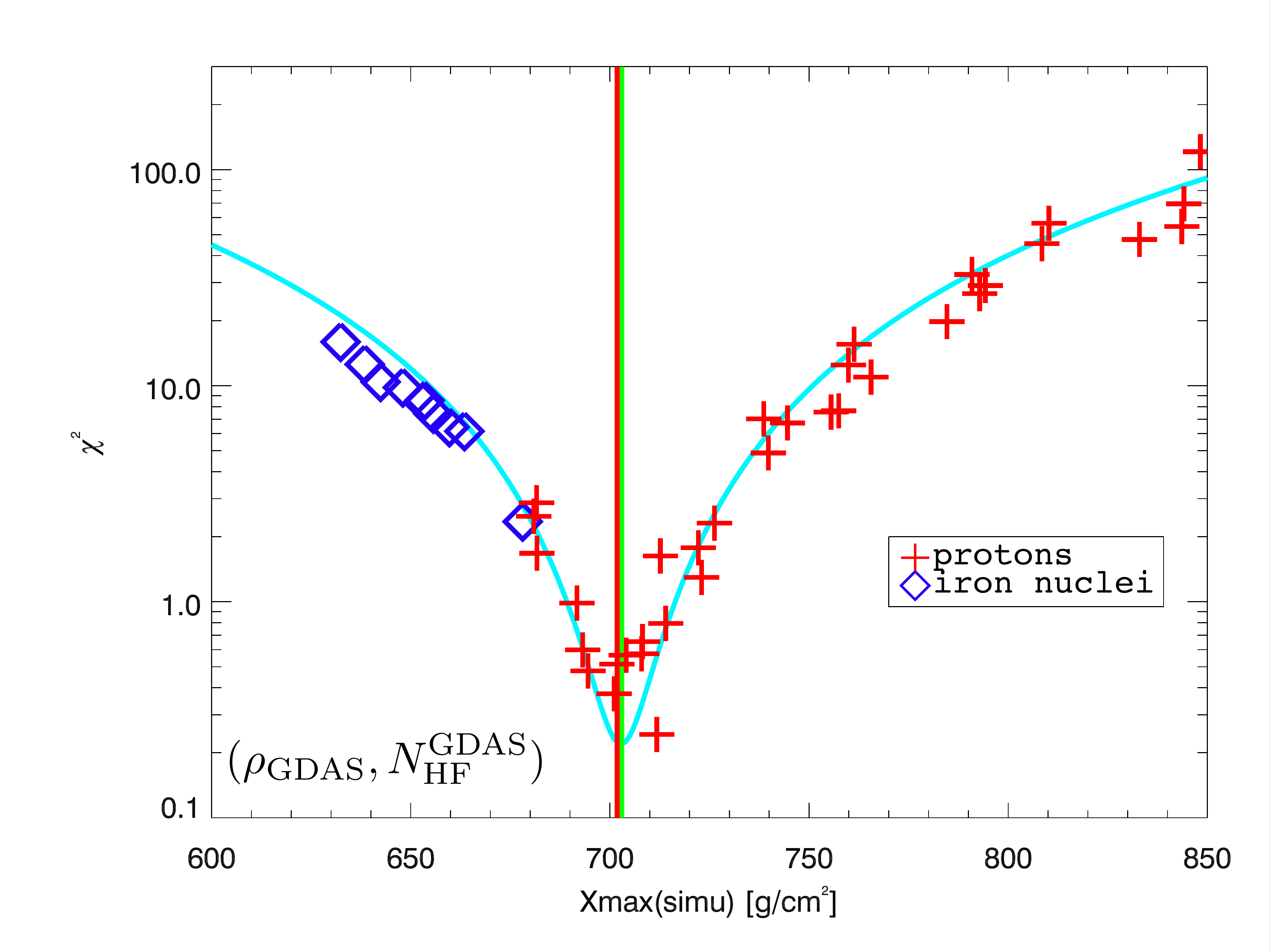}
  \includegraphics[width=9cm,height=5.2cm]{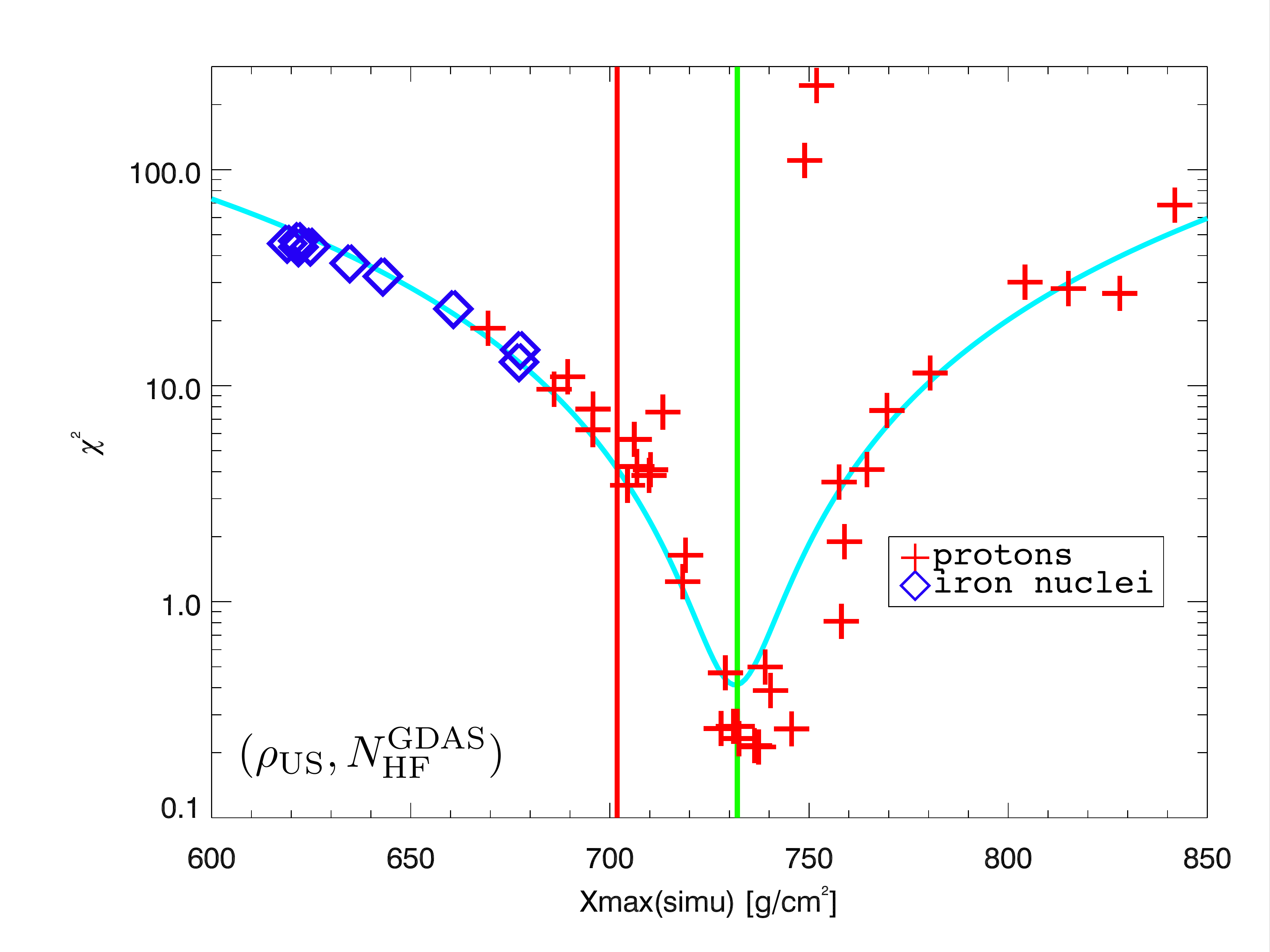}
   \caption{Value of the $\chi^2$ test as a function of the \xmax\ depths of simulated showers using $(\rho_\text{US},\nusgd)$ (top), $(\rho_\text{GDAS},\ngdashf)$ (middle) and $(\rho_\text{US},\ngdashf)$ (bottom). The red and green vertical lines correspond to the true \xmax\ of the test event and the reconstructed \xmax\ respectively.}
\label{fig:xmax}
  \end{center}
\end{figure}

The agreement between each simulated LDF to the test event is quantified through a $\chi^2$ test on the overall shape of the full 2D LDF (see~\cite{gatearena2016}). The 2D LDF is calculated as the maximum of the quadratic sum of the three polarizations. The results are shown in \fig~\ref{fig:xmax} for the three simulated sets as a function of the \xmax\ values of the simulated showers.

\tab~\ref{xmaxdmax} summarizes the results shown in \fig~\ref{fig:xmax}. We also give in this table the distance between the shower maximum and the shower core associated to the reconstructed \xmax.
\begin{table}[!ht]
\begin{center}
\begin{tabular}{|c|c|c|c|c|}
\hline
      & true value & $\ngdashf$ + $\rho_\text{GDAS}$& $\ngdashf$ + $\rho_\text{US}$ & $\nusgd$ + $\rho_\text{US}$\\
    \hline
     \xmax\ [$\gcm$] & 702 & 703 $\pm$ 7& 732 $\pm$ 8& 738 $\pm$14 \\
    \hline
    $D_\text{max}$ [m] & 4412 & 4408 $\pm$ 84 & 4396 $\pm$ 96 & 4323 $\pm$ 171 \\
    \hline
\end{tabular}
\end{center}
\caption{Reconstructed \xmax\ values for different atmospheric profiles and the corresponding geometrical distance $D_\text{max}$.}
\label{xmaxdmax}
\end{table}

We see that when using the correct description of the air density and air index values, we reconstruct with no bias the correct \xmax\ together with the correct distance to the shower maximum. With the US standard model for air density and the correct air index values ($\ngdashf$), the distance to the shower maximum is correctly estimated ($4396$~m instead of $4412$~m) but the \xmax/distance conversion is not satisfactory. This was expected as the overall shape of the 2D LDF is governed by the distance to the shower maximum which is strongly related to the air index model. Finally, taking a bad air density model (US standard) and a bad air index model leads to large discrepancies in both the \xmax\ and the distance to the shower maximum.
   \begin{figure}[!ht]
 \begin{center}
 \includegraphics[scale=0.3]{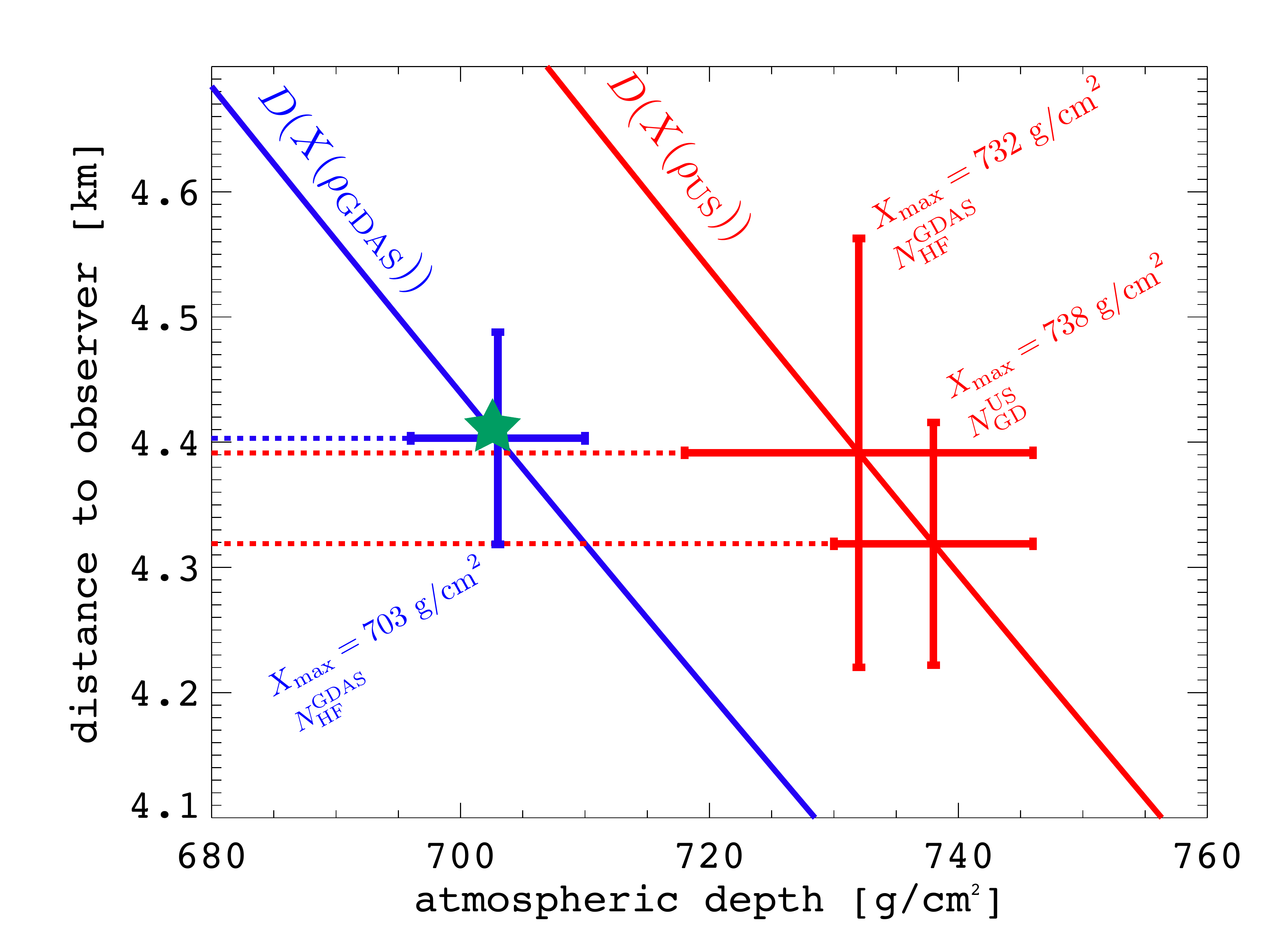}
   \caption{Geometric distance $D$ to the observer as a function of the atmospheric depth $X$ using $\rho_\text{US}$ (plain red curve) and $\rho_\text{GDAS}$ (plain blue curve). The two red points correspond to the reconstructions obtained with $(\rho_\text{US},\nusgd)$, $(\rho_\text{US},\ngdashf)$; the blue point corresponds to $(\rho_\text{GDAS},\ngdashf)$. The green star corresponds to the true value of the reference shower.}
\label{fig:distance}
  \end{center}
\end{figure}
This is also shown in \fig~\ref{fig:distance} where we display the relation between the distance between the shower core and the shower maximum as a function of the atmospheric depth, for the air density models $\rho_\text{GDAS}$ in blue and $\rho_\text{GDAS}$ in red.


\subsection{Set of reconstructed events}
\label{subsec:set}

To study the influence of both the air density $\rho$ and air refractivity $N$ on the \xmax\ reconstruction, we ran simulations corresponding to 6 different combinations of $(\rho,N)$.
Each set is composed of 10 iron showers and 40 proton showers, with random first interaction depth and \xmax, having the same arrival direction than our test event ($\theta=30^\circ, \phi=90^\circ$).

Set \#4 is taken as reference since it uses the most refined descriptions of both $\rho$ and $N$ on March 18, 2014 at noon. Each shower from  set \#4 is reconstructed using the 50 showers of the sets \#1, \#2, \#3, \#5, \#6 and the 49 other showers of set \#4.
The quality of the \xmax\ reconstruction (in $[30;80]$~MHz) is presented in \tab~\ref{tabxmax}.



\begin{table}[!ht]
\begin{center}
\begin{tabular}{|c|c|c|c|}
    \hline
     data set & air density& air index & $\Delta X_{30^\circ} [\gcm]$\\
    \hline
    \#1&US Std. & US Std. + GD  &  34.1 $\pm$ 8.9  \\
    \hline
    \#2&GDAS & US Std. + GD & 5.7 $\pm$ 5.4 \\
    \hline
    \#3&GDAS & GDAS + GD &  4.6 $\pm$ 3.6 \\
    \hline
   \#4&GDAS & GDAS + HF &   0.1$\pm$ 2.4 \\
    \hline
     \#5&GDAS & GDAS + HF (N+10$\%$) &  2.9 $\pm$ 4.8 \\
    \hline
     \#6&GDAS & GDAS + HF (N+20$\%$)  & 9.3 $\pm$ 16.4 \\
    \hline
\end{tabular}
\end{center}
\caption{Quality of the \xmax\ reconstruction for 6 different combinations of air density and air index. The $\Delta X$ column presents the mean difference with the true value and the 1$\sigma$ deviation.}\label{tabxmax}
\end{table}

The distributions of the differences are gaussian and the displayed values are the mean values and their standard deviations. Using set \#1, the mean difference is 34~$\gcm$, in agreement with the example presented in section~\ref{subsec:oneevent}. With set \#2, the differences are much smaller, because it uses the same GDAS air density profile than the tested events. With set \#3 the air index profile is calculated with the Gladstone and Dale law but with the air density profile from the GDAS data and the result is a bit better than set \#2. With set \#4 we use the most precise description by replacing the Gladstone and Dale law with the high frequency law. In this configuration the mean difference is compatible with zero ($0.1\pm 2.4~\gcm$), which also proves the self-consistency of the method. Sets \#5 and \#6 have been simulated with the same conditions as set \#4 but the refractivity $N$ has been artificially increased by 10\% and 20\%, respectively. The corresponding shift of the reconstructed \xmax\ values are 3~$\gcm$ and 10~$\gcm$. These results are in agreement with~\cite{papierholandais} where reconstructed \xmax\ values are shifted from 3.5 to 11~$\gcm$ for showers with zenith angles in the range $15^\circ-55^\circ$ in $[30;80]$~MHz for relative variation of $N$ of 4\%.

We also checked the influence of the zenith angle on the \xmax\ reconstruction. We repeated the same procedure with zenith angles $50^\circ, 55^\circ$~and $60^\circ$. We present the reconstruction in \tab~\ref{tabxmaxSELFAS3}.
\begin{table}[!ht]
\begin{center}
\resizebox{\textwidth}{!}{
\begin{tabular}{|c|c|c|c|c|c|c|}
    \hline
     data set & density& air index& $\Delta X_{30^\circ}$  & $\Delta X_{50^\circ}$ & $\Delta X_{55^\circ}$ & $\Delta X_{60^\circ}$ \\
    \hline
    \#1&US Std. & US Std. + GD  & $34.1 \pm 8.9$ & 51.1 $\pm$ 11.9 & 69.1 $\pm$ 9.2  & 108.1 $\pm$ 10.5 \\
    \hline
    \#2&GDAS & US Std. + GD & 5.7 $\pm$ 5.5 & 17.7 $\pm$ 9.5 & 21.0 $\pm$ 5.4 &  31.9 $\pm$ 10.9 \\
    \hline
    \#3&GDAS & GDAS + GD & 4.6 $\pm$ 3.6 & 9.1 $\pm$ 6.4 &  7.1 $\pm$ 4.6 &  4.7 $\pm$ 9.9  \\
    \hline
   \#4&GDAS & GDAS +HF &  0.1 $\pm$ 2.4 & 0.6 $\pm$ 5.4 &  0.9 $\pm$ 8.1 &  2.1 $\pm$ 10.0 \\
    \hline
     \#5&GDAS & GDAS + HF (N+10\%) & 2.9 $\pm$ 4.8 & 3.7 $\pm$ 6.3 &  4.0 $\pm$ 6.9 &  2.7 $\pm$ 12.7 \\
    \hline
     \#6&GDAS & GDAS + HF (N+20\%)  & 9.3 $\pm$ 16.4 & 9.5 $\pm$ 6.5 &  7.2 $\pm$ 9.9 &  3.9 $\pm$ 13.5 \\
    \hline
\end{tabular}}
\end{center}
\caption{Same as \tab~\ref{tabxmax} for three aditionnal zenith angles.}\label{tabxmaxSELFAS3}
\end{table}

These results tell us that the choice of the atmospherical model for the air density is the dominant factor in the quality of the \xmax\ reconstruction. Then, the air index model also has an influence on this quality, at a lower level though. In all cases, errors and bias increase with the zenith angle. Assuming the correct air index profile, the true geometrical distance to \xmax\ will be successfully reconstructed. The discrepancies arise from a bad conversion of this distance to its equivalent in traversed atmospheric depth, but these discrepancies, induced by the air density model, can be corrected successfully after the reconstruction, as presented in \cite{2016Natur.531...70B}. However, this is not the case for the errors induced by an incorrect air index profile modeling because the spatial and temporal structures of the electric field depend on the values of the air refractive index at the altitude of the emission maximum and at lower altitudes along the shower axis. As a consequence, the simulation has to be performed using the most refined model, namely, $\ngdashf$. 

\section{Conclusion}
In this work, we have studied the influence of the description of the atmosphere on the electric field emitted by air showers and its effect on the reconstruction of the properties of the primary cosmic ray using the radio technique. 

In order to reach the required accuracy to be a competitive technique, we need to describe the atmosphere in a very precise way. With this objective, we have demonstrated the need to use a spherical geometry for the Earth and its atmosphere: the flat approximation leads to systematic errors larger than $10~\gcm$ for zenith angles above~$60^\circ$.

After that, we have used the Global Data Assimilation System (GDAS), which provides information
on the atmospheric pressure, temperature and humidity for a range of altitudes every three hours.
These three quantities allow us to know the density of the atmosphere and its refractive index,
both of which are crucial for a correct simulation of the development of an extensive air shower
and the calculation of the electric field it produces. Since the data provided by the GDAS are
available up to $26$~km of altitude, our atmospheric model is a mixture of GDAS data below
$26$~km (which is the most important region for the development of air showers), and the
usual US Standard atmosphere above $26$~km.
The atmospheric refractivity has been calculated with two formulas: the usual Gladstone-Dale (GD)
formula, that does not take humidity into account, and a high-frequency (HF) 
formula that is more suited for
radio frequencies (MHz-GHz) and takes the relative humidity as an input. The differences in
refractivity between using a US Standard atmosphere coupled with the GD formula on one side
and the GDAS atmospheric data with the HF formula are of $~15\%$ on average at $1$~km
of altitude, and it can reach up to $35\%$ in the lowest layers of the atmosphere.

We have studied the influence of the refractivity on the time traces and
the lateral distribution function (LDF)
of the electric field produced by air showers. When considering the $[20;80]$~MHz band,
differences in refractivity up to $20\%$ result in a relatively small difference in the amplitude of the electric field and hence the LDF, indicating that
at these frequencies, an accurate knowledge of the refractivity is not the most important
factor for reconstructing the properties of the primary cosmic ray. However these small differences in the LDF vary as a function of the axis distance (+2\% to +8\% when increasing the refractivity by 20\%) leading to a shift in the reconstructed \xmax\ value. Moreover, when inspecting
the $[120;250]$~MHz band, the shape of the time traces for the electric fields and the
LDF on the ground change appreciably with the refractivity, making the reconstruction
at high frequency more dependent on the correct knowledge of the atmospheric refractivity.
In turn, if we can provide a precise refractivity, the $[120;250]$~MHz band presents the
advantage that the electric field footprint on the ground varies dramatically with the
shower maximum. In particular, the Cherenkov ring is clearly visible at these frequencies
and can help us discriminating the position of the shower maximum.

Finally, we have compared the performance in the reconstruction of the shower maximum
with the several atmospheric densities (US Standard and GDAS) and refractivities
(US Standard coupled with GD, GDAS coupled with GD and GDAS with the HF formula
with relative humidity) available. We have used test events simulated with the GDAS
density and HF refractive index, in order to quantify the error induced with the US
Standard atmosphere and the GD formula if we assume the GDAS data are closer to
the actual atmosphere. We have found that the most important parameter for the
reconstruction of the shower maximum is the air density, since even if we correctly
reconstruct the altitude of the shower maximum, an incorrect air density will bias
the atmospheric depth of the \xmax. The bias induced with a US Standard air density
lies around $\sim 30$ g/cm$^2$ for $30^\circ$ showers and $\sim 100$ g/cm$^2$
for $60^\circ$ showers. The bias induced by the refractivity calculated with the US Standard
atmosphere and the GD index ranges from $\sim 5$ g/cm$^2$ for $30^\circ$ showers
to $\sim 32$ g/cm$^2$ for $60^\circ$ showers. These biases are not negligible and
indicate the need for a correct description of the atmospheric properties. The theoretical
accuracy of the method, using the GDAS data and without taking into account uncertainties in the modelling of the electric field of the shower or the atmospheric parameters, is
$\sim 2.4$ g/cm$^2$ for $30^\circ$ showers and $\sim 10$ g/cm$^2$ for
$60^\circ$ showers. These accuracies constitute a theoretical limit for the precision
of the \xmax ~reconstruction using the method discussed in this paper.

To sum up with, the results of this paper indicate that a description of the atmosphere using the
US Standard model paired with the GD formula cause non-negligible biases when reconstructing
the \xmax, and therefore an alternative description is needed. The most complete description
of the atmosphere publically available is the GDAS data, from which we can trivially calculate
the properties of the atmosphere relevant for the simulation of the electric field produced by
air showers. In doing so, we guarantee the minimum possible bias in the simulation of
the electric field and the reconstruction of the shower maximum. Currently, 
the only way of improving this method is to directly measure the atmospheric 
properties for a given experiment \emph{in situ}.

\section*{Acknowledgement}
We would like to thank Martin Will, who kindly explained the scripts to access the GDAS data.

\section*{Appendix}
 \label{sec:Appendix}

At an altitude~$z$ and a latitude~$\phi$, the geopotential height is defined as:
\begin{equation}
Z_g(z,\phi)=\frac{1}{g_0} \int_{0}^{z} g(z',\phi) \, \mathrm{d}z'\label{zgeo}
\end{equation}
Where $g_0$ and $g(\phi,z)$ are respectively the gravitational acceleration at mean sea level and corrected for altitude $z$, latitude $\phi$ and Earth rotation. The function $g(\phi,z)$ can be estimated by the following relation known as International Gravity Formula~1967 with the free-air correction:

\begin{gather*}
g(z,\phi)=\Lambda(\phi) g_0 -C\,z\\
\text{with}~\Lambda(\phi)=1+A\sin^2(\phi)-B\sin^2(2\phi)\\
\text{and}~A=0.005 302 4,~B=5.8\times 10^{-6},~C=3.086\times 10^{-6}~\text{s}^{-2}.
\end{gather*}

We compute the altitude above sea level by solving \eq~\ref{zgeo}; the solution for $z$ at a given latitude $\phi$ and geopotential height $Z_g(z)$ is:

\begin{align*}
z(Z_g,\phi)~~ =~~ & \frac{g_0}{C}\left(\Lambda(\phi)-\sqrt{\Lambda^2(\phi)-\frac{2C\,Z_g}{g_0}}\right)\\
\end{align*}

\bibliographystyle{unsrt}
\bibliography{gdas}

\end{document}